\documentclass[lettersize,journal]{IEEEtran}
\usepackage{amsmath,amsfonts}
\usepackage{algorithmic}
\usepackage{algorithm}
\usepackage{array}
\usepackage[caption=false,font=normalsize,labelfont=sf,textfont=sf]{subfig}
\usepackage{textcomp}
\usepackage{stfloats}
\usepackage{color}
\usepackage{url}
\usepackage{verbatim}
\usepackage{graphicx}
\usepackage{cite}
\usepackage{booktabs}
\usepackage{bbding}
\usepackage{threeparttable}
\usepackage{wasysym}
\begin{document}

\title{Adaptive Probabilistic Constellation Shaping\\ based on Enumerative Sphere Shaping\\ for FSO Channel with Turbulence and Pointing Errors}

\author{Jingtian LIU\textsuperscript{(1)}, Bin CHEN\textsuperscript{(2)}, Senior Member, IEEE, Xiongwei YANG\textsuperscript{(1)}, Yi WEI\textsuperscript{(1)},\\ Jianjun YU\textsuperscript{(3)}, Fellow, IEEE, and Feng ZHAO\textsuperscript{(1)}
\thanks{These authors contributed equally: Jingtian LIU, Bin CHEN. E-mail of corresponding author: jingtian.liu@xupt.edu.cn, hfengzhao@xupt.edu.cn.}
\thanks{\textsuperscript{(1)} School of Electronic Engineering, Xi’an University of Posts and Telecommunications, Xi'an, 710121, China;~~\textsuperscript{(2)} School of Computer Science and Information Engineering, Hefei University of Technology, Hefei, 230009, China;~~\textsuperscript{(3)} School of Information Science and Technology, Fudan University, Shanghai, 200433, China.}}

\markboth{Journal of \LaTeX\ Class Files,~Vol.~14, No.~8, August~2021}%
{Shell \MakeLowercase{\textit{et al.}}: A Sample Article Using IEEEtran.cls for IEEE Journals}

\IEEEpubid{}

\maketitle

\begin{abstract}

Free-space optical (FSO) transmission enables fast, secure, and efficient next-generation communications with abundant spectrum resources. However, atmospheric turbulence, pointing errors, path loss, and atmospheric loss induce random attenuation, challenging link reliability. Adaptive coded modulation technology enhances spectrum utilization and reliability. We propose an adaptive probabilistic constellation shaping (A-PCS) coherent system utilizing enumerative sphere shaping (ESS) for distribution matcher (DM). With PCS-64QAM, the system achieves continuous rate control from conventional QPSK-equivalent to 64QAM spectral efficiency, providing quasi-continuous control with granularities of approximately $0.05$~bits/4D for spectral efficiency and $0.1$~dB for the post-FEC SNR threshold, and a maximum control depth of $12.5$~dB. Leveraging ESS for efficient sequence utilization, it offers higher spectral efficiency and finer control granularity than constant composition distribution matcher (CCDM)-based A-PCS systems. We further model and analyze the FSO channel, presenting calculations and comparisons of outage probability and ergodic capacity under varying turbulence intensities and pointing errors. Results demonstrate 99.999~\% reliability at maximum $\sigma_\mathrm{R}^2 = 1.02$ and $\sigma_\mathrm{s} = 0.51~\mathrm{m}$, meeting requirements under severe turbulence and large pointing errors. {Furthermore, under non-ideal delayed channel state information (CSI) feedback conditions, the system adapts to varying turbulence coherence times and feedback delays, with results showing that finer modulation granularity (provided by A-PCS-ESS) enhances immunity to feedback delay, maintaining a performance advantage over conventional adaptive schemes across a range of channel environments and delay values.}

\end{abstract}


\section{Introduction}

The rapid advancement of global digitalization has accelerated the development of sixth-generation (6G) mobile communication systems. In addition to achieving comprehensive terrestrial coverage, 6G requires the establishment of a robust integrated communication network spanning air, space, and sea domains~\cite{akyildiz2022terahertz,nie2021channel,han2025joint}. In free-space wireless communications, microwave-based systems offer a mature and reliable solution; however, their limited spectral resources in the GHz band constrain high-capacity data transmission, failing to meet the escalating demand for bandwidth in next-generation networks. In contrast, FSO communication leverages abundant spectrum resources at higher frequencies, offering significant advantages such as broad bandwidth, long transmission distance, and enhanced security~\cite{dhasarathan2020development,khalighi2014survey}. Moreover, FSO systems can interface seamlessly with existing fiber-optic infrastructures, supporting both line-of-sight and non-line-of-sight transmission modes.

Nevertheless, FSO signal propagation through the atmosphere is susceptible to various channel impairments. Variations in temperature and pressure along the transmission path induce refractive index inhomogeneities, leading to atmospheric turbulence. When combined with pointing errors caused by random vibrations at transceivers, these effects result in random fading of the received signal power~\cite{farid2007outage}. Adverse weather conditions-such as fog, haze, and snow-further attenuate the optical signal through photon absorption, causing severe degradation~\cite{prokes2009atmospheric}. Such channel characteristics pose substantial challenges to the reliability of FSO links.

To enhance link reliability, adaptive coded modulation schemes have been proposed~\cite{goldsmith1998adaptive, djordjevic2010adaptive, chatzidiamantis2011adaptive,  farid2009channel, nouri2022adaptive}, wherein modulation formats or error-correcting codes are dynamically selected according to channel conditions. For instance, an adaptive strategy based on received signal-to-noise ratio (SNR) is introduced in~\cite{mai2014adaptive}, switching among five modulation schemes from on-off keying (OOK) to $32$-ary pulse amplitude modulation (PAM) according to predefined SNR thresholds. When channel conditions deteriorate beyond the capability of OOK, the system switches to an radio frequency (RF) link employing modulations from binary phase shift keying (BPSK) to $32$-ary quadrature amplitude modulation (QAM). By accounting for atmospheric turbulence in FSO and multi-path effects in RF, the authors demonstrated significant gains in spectral efficiency and outage performance. Similarly, in~\cite{wang2019optimizing},low-density parity-check (LDPC) codes are employed, with code rates ranging from $\frac{1}{4}$ to $\frac{9}{10}$ and modulation orders from QPSK to $32$APSK, achieving $28$ adaptive switching levels. Combined with deep learning-based channel prediction, this approach further improved spectral efficiency and resilience against deep fading. Although increasing the number of modulation levels improves granularity and spectral utilization, the inherent spectral efficiency gap between discrete modulation formats limits the minimum control granularity to approximately 0.4~bits per four-dimensional symbol (bits/4D). 

To overcome this limitation, adaptive probabilistic constellation shaping (A-PCS) technology has emerged~\cite{buchali2015rate, wang2025adaptive, bocherer2015bandwidth}. By adjusting the probability of constellation points according to the Maxwell-Boltzmann (MB) distribution, PCS can achieve near-continuous adjustment of spectral efficiency while approaching the Shannon capacity limit. For intensity modulation direct detection (IM/DD) systems, A-PCS technology based on M-PAM and CCDM has been studied and evaluated under different atmospheric turbulence conditions~\cite{elzanaty2020adaptive}, initially demonstrating the superior performance of A-PCS compared to uniform signaling. For coherent systems, recent research has proposed and validated an A-PCS technique based on QAM and CCDM~\cite{guiomar2020adaptive, brandao2022cooperative}. By designing a time-window averaging algorithm for CSI prediction and introducing control redundancy, their experiments demonstrate that the system exhibits high transmission reliability and a significant overall throughput improvement. 

\begin{table*}[t]
    \centering 
    \caption{{Adaptive coded modulation schemes for FSO communication systems relevant for this work.}}
    \label{Tab:1}
    \begin{tabular}{cccccccc} 
        \toprule[1.5pt] 
        \textbf{Ref.} & \textbf{System} & \textbf{Mod. Format} & \textbf{Performance metric} & \textbf{Turbulence} & \textbf{Pointing errors} & \textbf{Turbulent coherence} & \textbf{CSI feedback}\\
        \midrule[0.75pt] 
        \cite{djordjevic2010adaptive} & IM/DD & A-PAM & Capacity/BER & \Checkmark & \XSolid & \XSolid & Ideal\\
        \cite{chatzidiamantis2011adaptive} & IM/DD & A-S-PSK & Capacity/BER & \Checkmark & \XSolid & \XSolid & Ideal\\
        \cite{nouri2022adaptive} & IM/DD & A-PAM+A-PPM & Capacity & \Checkmark & \XSolid & \XSolid & Ideal\\
        \cite{mai2014adaptive} & IM/DD & A-PAM & FER/$P_{\text{out}}$ & \Checkmark & \XSolid & \XSolid & Ideal \\
       \cite{wang2019optimizing} & CM/CD & A-PSK & Capacity & \XSolid & \XSolid & \XSolid & Ideal\\
       \cite{farid2009channel} & IM/DD & A-PCS-MLC-MSD \RIGHTcircle & MI/$P_{\text{out}}$ & \Checkmark & \XSolid & \XSolid & Ideal\\
       \cite{elzanaty2020adaptive} & IM/DD & A-PCS-PAM-CCDM \CIRCLE & Ergo. Capacity/FER & \Checkmark & \XSolid &  \XSolid & Ideal\\
       \cite{guiomar2020adaptive, brandao2022cooperative} & CM/CD & A-PCS-QAM-CCDM \CIRCLE & Capacity/NGMI &  \Checkmark & \XSolid & \XSolid & Ideal\\
       \textbf{This work} & CM/CD & A-PCS-QAM-ESS \Circle & Ergo. Capacity/GMI/$P_{\text{out}}$ & \Checkmark & \Checkmark & \Checkmark & Ideal+Delayed \\
        \bottomrule[1.5pt] 
    \end{tabular}
    \begin{tablenotes}
        \footnotesize
        \item {CM/CD: coherent modulation/coherent detection; A-S-PSK: adaptive subcarrier PSK; MLC: multi-level coding; MSD: multistage decoding; PPM: pulse position modulation; FER: frame error rate; Ergo.: ergodic; \Circle: low complexity coding/decoding; \CIRCLE: high complexity coding/decoding; \RIGHTcircle: low complexity coding, high complexity decoding.}
      \end{tablenotes}
\end{table*}

{Despite these advances, the design of A-PCS still faces several critical challenges, as highlighted by the comparative summary in Tab.~\ref{Tab:1}. First, existing studies uniformly assume an infinitely long block length for the DM, thereby neglecting the implementation complexity and rate loss that arise from finite-length DM in practical systems. Second, the channel impairments considered in prior work remain incomplete. Although most references in Tab.~\ref{Tab:1} incorporate atmospheric turbulence, they consistently ignore the impact of pointing errors and turbulent coherence. Moreover, all existing schemes rely on ideal (i.e., delay-free) CSI feedback, overlooking the significant degradation caused by feedback delays in realistic FSO channels. This incomplete modeling leads to inaccurate evaluations of outage probability and ergodic capacity under practical operating conditions. Therefore, a comprehensive analytical framework that jointly accounts for finite-length DM, realistic channel impairments (turbulence, pointing errors, and turbulent coherence), and non-ideal CSI feedback is urgently needed to accurately assess and optimize A-PCS performance in dynamic FSO environments.}

{This paper proposes an adaptive coding and modulation scheme with fine granularity, high spectral efficiency, and a wide dynamic control range. A dedicated analytical model is established for the outage probability and ergodic capacity of the A-PCS system, explicitly accounting for the finite block length effects. We adopt a Gamma-Gamma distribution model for atmospheric turbulence and a Rayleigh distribution model for pointing errors. An ESS algorithm is employed as the DM to probabilistically shape the $M$-QAM constellation, and a four-dimensional (4D) PCS encoder and decoder are constructed in conjunction with an LDPC encoder. We compare four different adaptive schemes under both ideal and non-ideal CSI feedback conditions: our proposed ESS-based A-PCS (A-PCS-ESS), a CCDM-based A-PCS (A-PCS-CCDM)~\cite{brandao2022cooperative}, a conventional discrete QAM-based adaptive scheme (A-QAM)~\cite{mai2014adaptive}, and the Shannon limit. The comparison is performed in terms of control range, control granularity, and net bit rate. Additionally, we analyze the outage probability and ergodic capacity of these schemes under varying intensities and coherence times of atmospheric turbulence and pointing errors. The main contributions of this paper are summarized as follows:}

\begin{itemize}
    \item {We propose a novel A-PCS architecture that utilizes an ESS for distribution matching in a coherent FSO system and demonstrates the complete control chain. This design features fine-grained control and a large control range, achieving quasi-continuous rate adaptive control from QPSK equivalent to 64QAM spectral efficiency.}

    \item {We incorporate the associated rate loss into the performance evaluation, considering the finite block length of the DM in practical deployments, providing a more realistic evaluation framework compared to works assuming infinite block lengths.} 
    
    \item {We derive the analytical expressions for the outage probability and ergodic capacity of the proposed system under a comprehensive channel model that takes into account the effects of atmospheric turbulence, pointing error, path loss, and atmospheric attenuation. Turbulent coherence time and feedback delay are also introduced to evaluate the system performance under non-ideal CSI feedback.}
\end{itemize}

The remainder of this paper is organized as follows: Section~\ref{System_Description} introduces the system model, PCS principles, and channel modeling; Section~\ref{The Proposed A-PCS Design with ESS} presents the system architecture and modeling of the proposed A-PCS scheme; Section~\ref{Simulation Results} evaluates the system performance under various channel conditions with ideal CSI feedback and non-ideal CSI feedback; and Section~\ref{Conclusion} summarizes the key findings and conclusions.

\section{System Description and Definitions}\label{System_Description}
\subsection{{System Model}}

\begin{figure*}[!ht]
  \centering
  \includegraphics[width=0.96\linewidth]{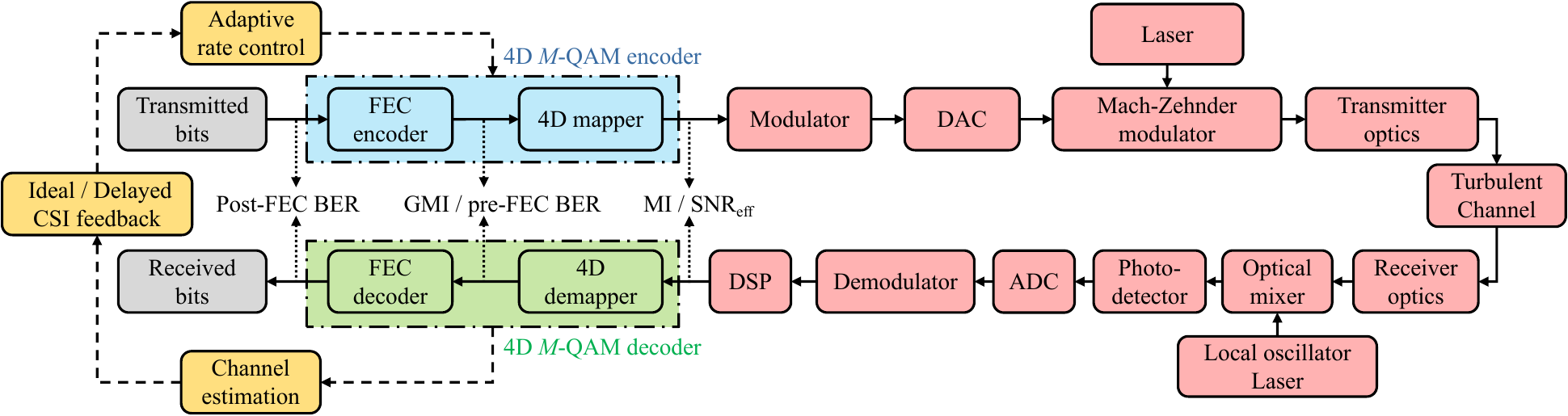}
  \caption{{The block diagram of adaptive coded modulation coherent FSO systems.}}
  \label{Fig. System_model}
\end{figure*}

{Fig.~\ref{Fig. System_model} illustrates an adaptive modulation communication system for FSO links based on CM/CD. At the transmitter, the input bits are first encoded by a forward error correction (FEC) encoder within the 4D $M$‑QAM encoder, which adds redundancy for protection. The encoded bits are then mapped to symbols in an $N$-dimensional signal space, selected from a constellation of $M = 2^m$ points. In a coherent FSO system, both orthogonal components of the electromagnetic field and two polarization states are exploited, each polarization state comprises two real dimensions: an in-phase (I) and a quadrature (Q) component. This naturally yields a 4D signal space ($N = 4$). For conventional two-dimensional modulation formats such as 64QAM, the symbols transmitted over the two polarization states are statistically independent and can be interpreted as polarization-multiplexed 4D symbols. After digital-to-analog conversion (DAC), the electrical signal is split into I and Q components, which drive a Mach–Zehnder modulator (MZM) to modulate two orthogonal optical carriers. The modulated optical signals then propagate through an atmospheric turbulent channel. At the receiver, the incoming beam is focused onto the aperture by the receiving optics and is combined with the local oscillator light in an optical mixer followed by a photodetector. The resulting electrical signal is subsequently processed by an analog-to-digital converter (ADC), a demodulator, and digital signal processing (DSP) to recover the transmitted symbols. In the 4D $M$‑QAM decoder, the recovered symbols are demapped into soft bits (e.g., log-likelihood ratios, LLRs) and passed to a soft-decision FEC decoder, which reconstructs the information bits. Using the received symbols and bits, CSI—including metrics such as received SNR, post/pre-FEC BER, and generalized mutual information (GMI) as described in Sec.~\ref{Sec. Performance Assessment Metrics},—is estimated. Through ideal or delayed CSI feedback (e.g., via global system for mobile communications (GSM)), the coding scheme is adapted in real time based on the observed channel conditions, thereby adjusting the transmission rate and enabling error-free communication.}

\subsection{Channel Model}
We consider a FSO communication system that employs coherent detection with either conventional or probabilistically shaped $M$-QAM. The input bit sequence, denoted as $\mathbf{B}^{k} = \{B_1, \ldots, B_k\}$, comprises independent and identically distributed bits following a Bernoulli distribution with parameter $1/2$. If the target symbol sequence is uniformly distributed, a binary reflected Gray code (BRGC) maps the bits directly to an $M$-QAM symbol sequence $\mathbf{S}^{n} = \{S_1, \ldots, S_n\}$, where $n = k / \log_2 M$. A forward error correction (FEC) code with rate $R_\text{C}$ is incorporated to ensure reliable transmission. Otherwise, PCS is applied: a DM generates a non-uniformly distributed symbol sequence using a binary labeling strategy, which integrates FEC coding. Further details on PCS are provided in Section~\ref{Probabilistic Constellation Shaping}.
The symbol sequence is subsequently filtered to produce an electrical signal, which modulates the optical source. After propagation through the channel, the optical beam is collected by a lens at the receiver and coherently mixed with a local oscillator for heterodyne detection, enabling coherent reception. In our simulations, laser phase noise and frequency offset are neglected, under the assumption that these impairments can be perfectly compensated via digital signal processing. Following filtering, the signal is down-sampled (due to the absence of phase noise). Finally, the electrical signal-to-noise ratio ($\mathrm{SNR_{elec}}$) is measured from the equalized constellations, and the generalized mutual information (GMI) is computed as described in Sec.~\ref{Sec. Performance Assessment Metrics}.

The FSO channel is subject to four primary impairments: atmospheric turbulence $h_\mathrm{a}$, atmospheric loss $h_\mathrm{l}$, pointing errors $h_\mathrm{p}$~\cite{farid2007outage}, and geometric spread. For short-range links, geometric spread is integrated into the pointing errors component as a geometric expansion term. These impairments, combined with ambient light noise and thermal noise-modeled accurately as additive white Gaussian noise (AWGN)~\cite{zhu2002free}-directly degrade the SNR at the receiver.
Based on the above effects, the received signal is conventionally modeled as~\cite{aman2025effective}:
\begin{equation}
y = (h \kappa P_\mathrm{x})^{\tau/2} x + n, \quad \text{where} \quad h = h_\mathrm{l} h_\mathrm{p} h_\mathrm{a}.
\label{eq:FSO Channel model}
\end{equation}
where $x$ represents the transmitted signal, $\tau$ is a detector-specific parameter ($\tau = 1$ for heterodyne detection and $\tau = 2$ for direct detection), $h_\mathrm{l}$ denotes the atmospheric loss (modeled via the Beer-Lambert law and assumed constant), $h_\mathrm{p}$ and $h_\mathrm{a}$ are random variables following Gamma-Gamma and Rayleigh distributions, respectively (as detailed below), $\kappa$ is the receiver's optical-to-electrical conversion efficiency, and $n$ is AWGN with variance $\sigma_\mathrm{n}^2$. The corresponding instantaneous SNR is given by:
\begin{equation}
\gamma_\mathrm{FSO}(h) = \frac{(h \kappa P_\mathrm{x})^\tau}{\sigma_\mathrm{n}^2},
\end{equation}
where $h$ is a random variable whose distribution is specified in Eq.~\eqref{eq:h}.

\subsubsection{Atmospheric loss}

The atmospheric attenuation coefficient characterizes the total loss due to absorption and scattering by atmospheric molecules and aerosols. At a wavelength of $1550$~nm, molecular absorption is negligible, with an approximate value of $0.01$~dB/km. The dominant factor in atmospheric attenuation is particle scattering, which encompasses molecular scattering, aerosol scattering, and scattering by hydrometeors such as haze, fog, snow, rain, and hail. This attenuation can be modeled using the exponential Beer-Lambert law as~\cite{prokes2009atmospheric}:
\begin{equation}
h_{\mathrm{l}}(\lambda,z) = \exp(-\sigma(\lambda)z),
\end{equation}
where $\sigma(\lambda)$ is the wavelength-dependent attenuation coefficient, which is a function of wavelength and the meteorological visibility range $V$. The meteorological visibility range is defined as the distance at which the transmittance at $550$ nm falls to a threshold of $\epsilon = 0.05$~\cite{prokes2009atmospheric}, {which directly characterizes the severity of scattering caused by fog, haze, rain, snow, and other hydrometeors}. (Note: Increased fog concentration is positively correlated with reduced turbulence intensity~\cite{kim2001comparison}.)

\subsubsection{Atmospheric turbulence}

Atmospheric turbulence arises from temperature and pressure gradients that induce refractive index inhomogeneities. These inhomogeneities form turbulent eddies that scatter optical beams based on the ratio of eddy size to beam width. When the eddy size is comparable to the beam width, focusing and defocusing effects cause random irradiance fluctuations at the receiver, known as scintillation. Scintillation, manifested as intensity fluctuations, is the dominant fading effect in FSO channels, degrading the SNR through deep fades~\cite{andrews1999theory}.
Several statistical models describe scintillation-induced fading: the log-normal distribution for weak-to-moderate turbulence~\cite{al2001mathematical}, the double Weibull distribution for moderate-to-strong turbulence~\cite{chatzidiamantis2010new}, the K-distribution for strong turbulence~\cite{samimi2012distribution}, and the Gamma-Gamma (GG) distribution for weak-to-strong turbulence~\cite{kaushal2017free}. Since our analysis covers a wide range of turbulence conditions from weak to strong, we employ the GG distribution, which agrees well with experimental measurements and models the fading as the product of large-scale and small-scale eddy effects.
The probability density function (PDF) of the GG distribution for the intensity $h_\mathrm{a}$ is given by~\cite{kaushal2017free}:
\begin{equation}
    f(h_\mathrm{a}) = \frac{2(\alpha\beta)^{\frac{\alpha + \beta}{2}}h_\mathrm{a}^{\frac{\alpha + \beta}{2}-1}}{ \Gamma(\alpha) \Gamma(\beta) } \text{K}_{\alpha-\beta}\left(2 \sqrt{\alpha \beta h_\mathrm{a}}\right), h_\mathrm{a} > 0
    \label{eq:ha}
\end{equation}
where $K_{a}(\cdot)$ is the modified Bessel function of the second kind of order $a$, $\Gamma(\cdot)$ is the gamma function, and $\alpha$ and $\beta$ represent the effective numbers of small-scale and large-scale eddies, respectively. These parameters are defined as:
\begin{equation}
\alpha = \left[ \exp\left( \frac{0.49\chi^{2}}{ (1 + 0.18d^2 + 0.5\chi^{12/5})^{7/6} } \right) - 1 \right]^{-1},
\label{eq:alpha}
\end{equation}
\begin{equation}
\beta = \left[ \exp\left( \frac{0.51\chi^{2} (1 + 0.69\chi^{12/5})^{-5/6} }{ (1 + 0.9d^2 + 0.62d^2\chi^{12/5})^{7/6} } \right) - 1 \right]^{-1},
\label{eq:beta}
\end{equation}
where $\chi^2 = 0.5 C_n^2 k^{7/6} z^{11/6}$ and $d = \sqrt{k a^2 / (4z)}$. Here, $k = 2\pi/\lambda$ is the optical wave number, $a$ is the diameter of the receiver aperture, $z$ is the link range in meters, and $C_n^2$ is the refractive index structure parameter, which depends on temperature, pressure, and humidity.
The turbulence severity is determined by the refractive index structure parameter, wave number, and link range, often quantified by the Rytov variance, $\sigma_\mathrm{R}^2$~\cite{andrews2001laser}:
\begin{equation}
\sigma_\mathrm{R}^2 = 1.23 C_n^2 k^{7/6} z^{11/6} = 2.46 \chi^2.
\label{eq:Rytov}
\end{equation}

\begin{figure}[!t]
\centering
\includegraphics[width=1\linewidth]{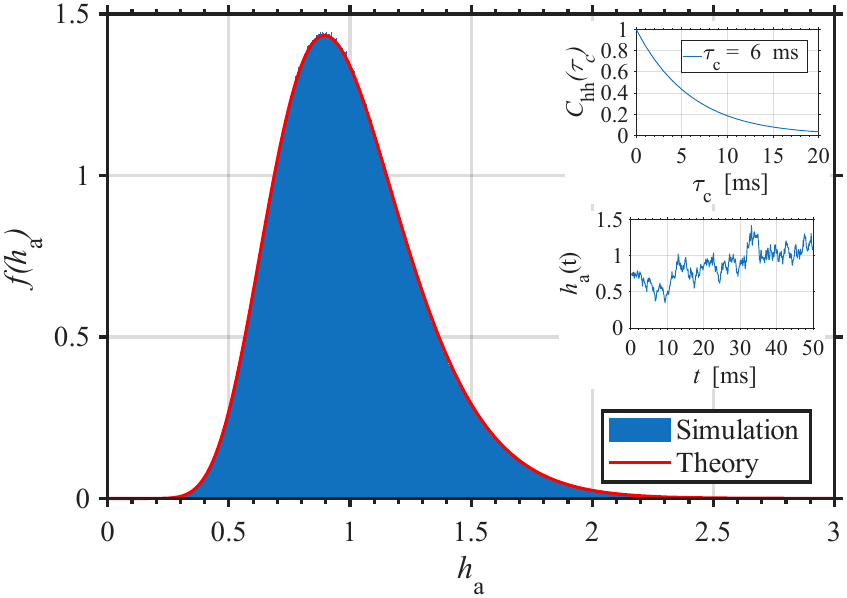}
\caption{{The empirical PDF of the generated GG process by SDE and its theoretical plots for $\sigma_\mathrm{R}^2 = 0.1$ with coherence time $\mathrm{\tau_\mathrm{c}}$~=~6~ms. The inset figures display the autocovariance $C_{\mathrm{hh}}$ of the generated GG process in the upper panel, and an example of time-correlated turbulence intensity in the lower panel.}}
\label{Fig. h_a_f_h_a}
\end{figure}

Thus, in our simulations, the turbulence strength is varied through $\sigma_\mathrm{R}^2$, and $h_\mathrm{a} = 1$ under turbulence-free conditions.
The fading process in atmospheric turbulence channels has a coherence time on the order of $10^{-3}$ to $10^{-2}$ seconds, which is much longer than typical symbol durations (around $10^{-9}$ seconds)~\cite{davidson2010channel,jurado2011computationally}. Therefore, the channel gain $h_\mathrm{a}$ remains approximately constant during each transmission. This slow fading makes interleaving techniques ineffective for averaging over fading states. {Consequently, such channels exhibit block-fading characteristics (also known as slow-fading or non-ergodic channels) and can be modeled using exponentially correlated processes via first-order stochastic differential equations (SDE)~\cite{bykhovsky2016simple} as shown in Fig.~\ref{Fig. h_a_f_h_a}.}

\subsubsection{Pointing Errors} \label{Sec: Pointing error}
In line-of-sight FSO communication links, the light beam propagating through the atmosphere undergoes geometric expansion, leading to a continuous increase in beam width. Since the received beam width typically exceeds the size of the receiving aperture, only a portion of the transmitted energy is captured, resulting in deterministic geometric expansion loss. This loss is minimized when the beam center aligns perfectly with the receiving aperture. However, random factors such as building vibrations induced by wind loads or thermal expansion can cause transmit beam jitter~\cite{arnon2003effects}, leading to misalignment between the received beam and the photodiode aperture. This misalignment prevents the beam focus from coinciding with the aperture focal point, degrading pointing accuracy and further attenuating the received signal strength. The fixed geometric spread is correlated with random pointing accuracy: a larger received beam width reduces the attenuation depth due to pointing errors but increases the geometric spread loss. These effects are commonly modeled jointly. For a Gaussian beam, the pointing errors loss $h_{\mathrm{p}}$ due to misalignment can be expressed in Gaussian form as:
\begin{equation}
h_{\mathrm{p}}(r; z) \approx A_0 \exp\left(-\frac{2r^2}{w^2_{z_{\mathrm{eq}}}}\right),
\end{equation}
where $r$ denotes the pointing error, defined as the radial distance between the center of the beam footprint and the center of the detector. By assuming that the elevation and horizontal sway displacements follow independent and identical Gaussian distributions~\cite{farid2007outage}, the PDF of $h_{\mathrm{p}}$ can be derived as:
\begin{equation}
f_{h_{\mathrm{p}}}(h_{\mathrm{p}}) = \frac{\gamma^2}{A_0^{\gamma^2}} h_{\mathrm{p}}^{\gamma^2 - 1}, \quad 0 \leq h_{\mathrm{p}} \leq A_0,
\label{eq:hp}
\end{equation}
where $\gamma = w_{z_{\mathrm{eq}}} / (2\sigma_{\mathrm{s}})$ represents the ratio of the equivalent beam radius $w_{z_{\mathrm{eq}}}$ at the receiver to the standard deviation of the pointing error displacement (jitter) $\sigma_{\mathrm{s}}$. The parameter $A_0$ corresponds to the fraction of collected power when the beam center coincides with the receiving aperture, accounting for inherent beam spreading effects. Thus, $h_{\mathrm{p}} = A_0$ in the absence of pointing error.
\subsubsection{Composite Channel Model}
Given the PDFs of the atmospheric turbulence gain $h_{\mathrm{a}}$ and the pointing errors gain $h_{\mathrm{p}}$, as provided in Eq.~\eqref{eq:ha} and Eq.~\eqref{eq:hp}, respectively, the PDF of the composite channel gain coefficient $h = h_{\mathrm{l}} h_{\mathrm{p}} h_{\mathrm{a}}$ can be expressed Eq.~(15) in~\cite{sandalidis2009optical}:
\begin{equation}
\begin{split}
    f_\mathrm{h}(h) = & \frac{\alpha \beta\gamma^2}{A_\mathrm{0}h_{\mathrm{l}}\Gamma(\alpha) \Gamma(\beta)} \\& \times  G_{1,3}^{3,0} \left[ \frac{\alpha \beta}{A_\mathrm{0}h_{\mathrm{l}}}h \,\middle|\begin{array}{c} \gamma^2 \\ \gamma^2-1, \alpha-1, \beta-1 \end{array}  \right],
    \label{eq:h}
\end{split}
\end{equation}
where the integration can be evaluated using numerical techniques.


\subsection{Probabilistic Constellation Shaping with Enumerative Sphere Shaping} \label{Probabilistic Constellation Shaping}
\begin{figure*}[!ht]
  \centering
  \includegraphics[width=1\linewidth]{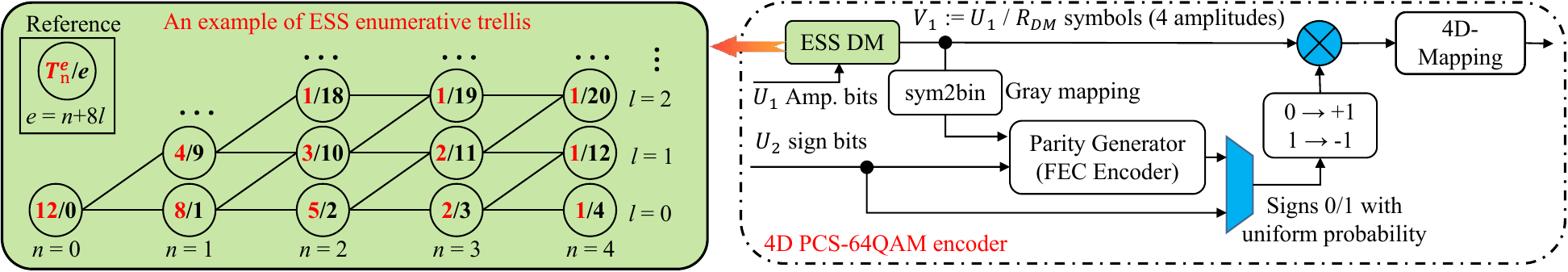}
  \caption{Block diagram of 4D PCS-64QAM encoder with FEC and ESS distribution matcher. An example of bounded-energy enumerative amplitude trellis for $\mathcal{A} = \{1, 3\}$, $N = 4, L = 3$, and $E_{\mathrm{max}} = 20$, is shown in the left image, which can achieve A-PCS encoding at arbitrary spectral efficiency by changing the energy boundary through setting the value of $L$. {The PCS decoder at the receiving end simply needs to first perform iterative decoding and error correction on the FEC codes, and then apply the ESS distribution matching decoder algorithm to decode the amplitude information in order to recover the original information~\cite{amari2019introducing}.}}
  \label{Fig. A_PCS_Encoder}
\end{figure*}
PCS aims to generate symbol sequences that adhere to a target probability distribution. As established in~\cite{kschischang1993optimal}, the MB distribution is the optimal symbol probability distribution for the AWGN channel under a fixed average energy constraint. A key component in achieving this distribution is the DM, which transforms uniformly distributed input symbol sequences into sequences conforming to the target distribution. These sequences emulate discrete memoryless sources and are designed to be invertible, ensuring accurate recovery of the original input symbols via a dematching process.

Let the input sequence consist of $k$ independent and identically distributed bits, denoted as $\mathbf{B^{k}} = \{B_1, \cdots, B_k\}$, each following a Bernoulli distribution with probability $\frac{1}{2}$. In the fixed-to-fixed DM process, a bijective mapping function $f$ maps $\mathbf{B^{k}}$ to a length-$N$ sequence $\mathbf{\tilde{A}^N} = f(\mathbf{B^{k}})$, where $\mathbf{\tilde{A}^N} \in A^N$. The output distribution $P_{\mathbf{\tilde{A}}}$ approximates the desired MB distribution $P_{\mathrm{A}}$
\begin{equation}
    P_{A}(a) =
    \begin{cases}
    K(\lambda)e^{-\lambda a^2}, & \text{for } a \in A, \\ 
    0, & \text{otherwise},
    \end{cases}
    \label{Eq. lambda}
\end{equation}
where $\lambda$ is the shaping parameter controlling the distribution's variance, and $K(\lambda)$ is a normalization constant.

This mapping is crucial in PCS for tailoring signal constellations to specific channel conditions, thereby improving system performance. The bijectivity of $f$ guarantees reversibility through $f^{-1}$, ensuring reliable data recovery in optical communication systems.

CCDM, introduced by Schulte and Böcherer~\cite{schulte2015constant}, generates sequences with a fixed symbol composition. However, CCDM suffers from significant rate loss at short block lengths and requires long sequences to achieve high shaping gains. In contrast, ESS~\cite{amari2019introducing,Gultekin20}, proposed by Amari et al. and extended by G\"ultekin et al., maps input bits onto a high-dimensional sphere to optimize energy efficiency. ESS exhibits lower rate loss than CCDM at the same shaping rate. For a block length of 108, ESS achieves only one-fifth the rate loss of CCDM. This advantage stems from ESS utilizing all energy levels below a fixed maximum energy. As the block length increases, the performance of ESS and CCDM converges.

Sphere coding constrains sequences to an $N$-dimensional sphere. The induced one-dimensional distribution converges to the MB distribution as $N \rightarrow \infty$. ESS employs bounded-energy amplitude sequences (up to $E_{\mathrm{max}}$) and uses lexicographic indexing to enumerate sequences. Graphically, ESS can be represented by a bounded-energy trellis where states correspond to energy levels and branches represent amplitudes. In this trellis, each sequence $\mathbf{a^{\textit{N}}} = (a_1, \cdots, a_N)$ corresponds to a unique path of $N$ branches. A branch from column $n-1$ to $n$ represents the amplitude $a_n$. Nodes in column $n$ indicate the cumulative energy up to dimension $n$
\begin{equation}
    E((a_1, \cdots, a_n)) = \sum_{i=1}^{n} a_i^2,
\end{equation}
where energy values are shown as black numerals in Fig.~\ref{Fig. A_PCS_Encoder}, paths start from the zero-energy node (bottom-left) and end at column $N=4$, the energy increments by $1$ or $9$ for $a_n = 1$ or $3$, respectively, terminal nodes have energies in $\{N, N+8, N+16, \ldots, E_{\mathrm{max}}\}$ for $\mathcal{A} = \{1, 3\}$, the energy at any node in column $n$ is given by
\begin{equation}
    e = n + 8l,
\end{equation}
where $l \in [0, L-1]$ and $L$ is the number of energy levels
\begin{equation}
    L = \left\lfloor \frac{E_{\mathrm{max}} - N}{8} \right\rfloor + 1,
    \label{Eq.L}
\end{equation}
where $ \left\lfloor \cdot  \right\rfloor$ is the floor operation, and $E_{\mathrm{max}}$ can be derived as
\begin{equation}
    E_{\mathrm{max}} = 8(L - 1) + N.
    \label{Eq.E_max}
\end{equation}

In Fig.~\ref{Fig. A_PCS_Encoder}, each node is identified by the pair $(n, e)$. The red numerals in the figure denote the number of paths to a node at $n = N$, which is computed recursively for $n = N-1, \ldots, 0$ and $0 \leq e \leq E_{\mathrm{max}}$ as
\begin{equation}
T_n^e = \sum_{a \in A} T_{n+1}^{e + a^2},
\end{equation}
with the initial condition $T_N^e = 1$. Only states with energy $e$ satisfying $e \in [n, E_{\mathrm{max}} + n - N]$ are considered in each column. The input block length can then be expressed as
\begin{equation}
k = \left\lfloor \log_2\left| T_0^0 \right| \right\rfloor \quad (\text{bits}),
\end{equation}
the rate of this ESS is defined as the ratio of the input bit length to the number of output amplitudes, and is given by
\begin{equation}
R_{\mathrm{DM}} = \frac{k}{N} \quad (\text{bits/amp.}),
\end{equation}
the amplitude distribution $P_A(a)$ over the set $\mathcal{A}$ is computed as follows (see Lemma~5.1 in \cite{gultekin2020enumerative})
\begin{equation}
P_{\mathrm{A}}(a) = \frac{T_{1}^{a^2}}{\sum_{b \in \mathcal{A}} T_{1}^{b^2}},
\end{equation}
the average energy of the sequences, denoted $E_{\mathrm{av}}$, is given by
\begin{equation}
E_{\text{av}} = N \sum_{a \in \mathcal{A}} P_A(a)a^2,
\end{equation}
hence, the rate loss is defined as the difference between the entropy of the MB distribution and the rate of the ESS, both evaluated at the same average energy
\begin{equation}
R_{\mathrm{loss}} = H(A_{\mathrm{MB}}) - R_{\mathrm{DM}},
\label{eq:rate_loss}
\end{equation}
where $A_{\mathrm{MB}}$ denotes a random variable following the MB distribution with average energy $E_{\mathrm{av}}$. The rate loss formula in Eq.~\eqref{eq:rate_loss} applies to both ESS and CCDM. As $N \rightarrow \infty$, ESS asymptotically achieves the MB distribution, and consequently $R_{\mathrm{loss}} \rightarrow 0$.

{\textbf{Example 1} \textit{Consider the ESS with blocklength $N = 4$ shown in Fig.~\ref{Fig. A_PCS_Encoder}. From node $(3,11)$, paths lead to either $(4,12)$ (corresponding to amplitude $1$) or $(4,20)$ (corresponding to amplitude $3$), so $T_3^{11} = 2$. From node $(1,1)$, we have $T_1^1 = 7$. Calculating in a similar way as above, we get the total number of sequences is $T_0^0 = 11$, which encodes $k = \lfloor \log_2 T_0^0 \rfloor = 3$ bits. The distribution matcher rate is $R_{\mathrm{DM}} = 0.75~\mathrm{bits/amp.}$. The resulting amplitude distribution gives $P_A(1) = \frac{7}{11}$ and $P_A(3) = \frac{4}{11}$, with an average energy $E_{\mathrm{av}} = 15.6364$. The rate loss is computed as $R_{\mathrm{loss}} = H(A_{\mathrm{MB}}) - R_{\mathrm{DM}} = 0.95 - 0.75 = 0.2~\mathrm{bits/amp.}$}.}



\subsection{PCS Performance Assessment Metrics} \label{Sec. Performance Assessment Metrics}
To address the complexity inherent in coded modulation for optical communication systems, bit-interleaved coded modulation (BICM) is widely adopted as an efficient approach~\cite{alvarado2015four}. BICM separates the coding and mapping processes, facilitating a more flexible and practical system design. In BICM systems, forward error correction (FEC) coded bits undergo interleaving, which disperses potential burst errors during transmission, thereby enhancing the robustness of the communication link.

After interleaving, the bits are grouped into blocks and mapped onto modulation symbols for transmission over the optical channel. At the receiver, the system demodulates these symbols to recover the original information. To support this, soft information-typically in the form of bit-wise a posteriori probabilities or log-likelihood ratios-is extracted from the demodulated symbols and passed to the decoder. This procedure is referred to as bit-metric decoding (BMD).

BMD is particularly effective in channels with non-uniform error characteristics, as it provides a reliability measure for each bit based on the received symbol. The flexibility of BICM, combined with the efficiency of BMD, makes it an attractive coding scheme for modern optical communication systems, offering a balance between performance and complexity.

BICM can be modeled as $m$ parallel, memoryless, and independent binary channels, where $m = \log_2(M)$ for an $M$-symbol constellation. The maximum achievable information rate of BICM is known as the generalized mutual information (GMI), defined as~\cite{bocherer2017achievable,cho2017normalized}:
\begin{equation}
\begin{split}
    \mathrm{GMI}_{\mathrm{BMD}} & (\gamma_{\mathrm{FSO}}) \approx H(X) - \\& \frac{1}{N} \sum_{k=1}^N \sum_{i=1}^m \log_2 \frac{\sum_{x \in A} q_{Y|X}(y_k|x)p_X(x)}{\sum_{x \in A_{b_{k,i}}} q_{Y|X}(y_k|x)p_X(x)},    
\end{split}
\end{equation}
Here, $q_{Y|X}$ denotes the auxiliary channel conditional probability, $y_k$ represents the $k$-th received symbol out of $N$ symbols in an AWGN channel with signal-to-noise ratio $\gamma_{\mathrm{FSO}}$, and $A_{b_{k,i}}$ is the set of 4D constellation symbols whose $i$-th bit equals $b_{k,i} \in \{0,1\}$. The final term, corresponding to channel loss, can be estimated via Monte Carlo methods. Although BICM incurs a capacity loss compared to coded modulation-constrained capacity, employing Gray mapping minimizes this loss, making it acceptable for practical applications.

Furthermore, in PCS, DM is used to optimize the transmitted symbol distribution, thereby improving system efficiency and performance. Considering finite-length DM is crucial, as practical implementations involve constrained DM lengths. Conventionally, GMI is estimated under the assumption of infinite-length DM, where rate loss is negligible. However, in real-world scenarios, finite DM length introduces a non-negligible rate loss that must be accounted for.

The GMI computation for finite-length one-dimensional DM involves evaluating the information content per amplitude shift keying (ASK) symbol for both infinite and finite DM lengths. This requires incorporating the rate loss due to finite DM into the GMI estimation, enabling a more accurate assessment of system performance. Thus, the GMI for finite-length DM in 4D $M$-QAM is defined as:
\begin{equation}
\label{eq:GMI_DM}
    \mathrm{GMI}_{\mathrm{4D}}(\gamma_{\mathrm{FSO}}) = \mathrm{GMI}_{\mathrm{BMD}}(\gamma_{\mathrm{FSO}}) - 4R_{\mathrm{loss}},
\end{equation}
where $R_{\mathrm{loss}}$ is introduced in Eq.~\eqref{eq:rate_loss}, with detailed derivations provided in the appendix of~\cite{Fehenberger19}. The factor of 4 accounts for the four dimensions in 4D~$M$-QAM ((two pairs of I and Q signals in two polarizations). Eq.~\eqref{eq:GMI_DM} is used in subsequent sections to compare GMI across DMs with varying rate losses. {We note that the post-FEC SNR thresholds used in this work are derived from GMI rather than direct FER simulations. However, for the DVB-S2 LDPC code employed (length 64,800, rate 5/6), the GMI has been shown to provide a tight bound, typically within $0.1$–$0.2$~dB of the BER waterfall region under AWGN~\cite{alvarado2015four, cho2017normalized}. Given our control granularity of $0.1$~dB and the additional SNR redundancy $\gamma_\rho$ in Sec.IV.B added for non-ideal feedback, the GMI-based threshold serves as a reliable and conservative estimate.}

\begin{figure*}[!ht]
  \centering
  \includegraphics[width=1\linewidth]{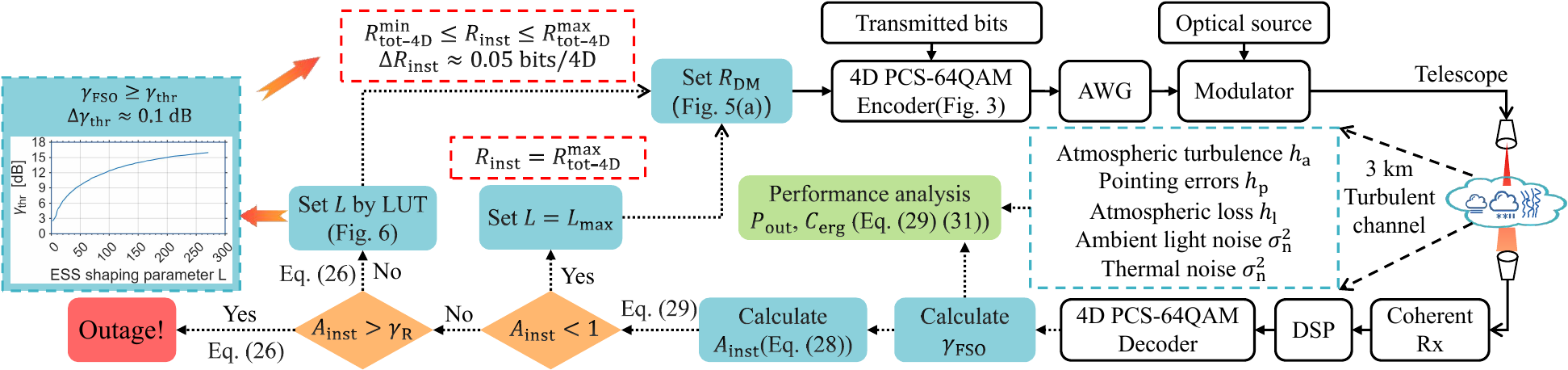}
  \caption{{Block diagram of adaptive 4D PCS-64QAM system architecture with FEC and ideal CSI feedback. The channel effects are shown in the middle.}}
  \label{Fig. System_FSO_PCS}
\end{figure*}

\section{The Proposed A-PCS Design with ESS} \label{The Proposed A-PCS Design with ESS}
This study investigates an A-PCS FSO communication system that utilizes a four-dimensional (4D) PCS-64QAM encoder at the transmitter, as illustrated in Fig.~\ref{Fig. System_FSO_PCS}. The system employs ESS distribution matching (DM) with a block length of $n = 108$, selected to balance complexity and rate loss. Additionally, it incorporates a LDPC code (length = 64,800 bits, code rate $R_\mathrm{C} = \frac{5}{6}$) compliant with the DVB-S2 standard.

\subsection{Implementation of A-PCS}
As depicted in Fig.~\ref{Fig. A_PCS_Encoder}, the input bit stream is partitioned into two components: $U_1$ amplitude bits and $U_2$ sign bits. The amplitude bits $U_1$ are fed into the ESS DM, which encodes them into amplitude symbols $V_1 = \frac{U_1}{R_\mathrm{DM}}$ for $M$-QAM modulation. For 64QAM, the amplitude levels are 1, 3, 5, and 7. Here, $R_\mathrm{DM}$ denotes the distribution matcher rate, determined by the ESS shaping parameter $L$. Subsequently, for each symbol, a 2-bit pattern is assigned via Gray mapping. These Gray-mapped bits, along with the uniformly distributed sign bits $U_2$, are input to the forward error correction (FEC) encoder operating at a rate of $R_\mathrm{C} = \frac{k}{n}$. The redundant parity bits ($n - k$) from the FEC encoder are multiplexed with $U_2$ to form the symbol signs based on the $(-1)^p$ mapping (i.e., $0 \mapsto +1$ and $1 \mapsto -1$). Finally, the symbols are mapped to dual-polarization 4D 64QAM for transmission. The total net bit rate of the 4D $\text{PCS-}M\text{-QAM}$ system is given by:

\begin{equation}
    R_\mathrm{tot\text{-}4D} = 4 + 4R_{\mathrm{DM}} - 2m(1 - R_\mathrm{C}),
    \label{Eq.R_total}
\end{equation}

where $0 < R_{\mathrm{DM}} \le \frac{m-2}{2}$ and $\frac{m-2}{m} \le R_\mathrm{C} < 1$. The minimum value of $R_{\mathrm{DM}}$ corresponds to the DM outputting a single amplitude level, equivalent to 4D QPSK. The maximum value of $R_{\mathrm{DM}}$ indicates no shaping is applied, resulting in conventional 4D $M$-QAM. The minimum value of $R_\mathrm{C}$ occurs when $U_2 = 0$, meaning all required sign bits can be generated solely from FEC encoding of the DM output; a lower FEC rate would produce redundant bits that cannot be accommodated. For 64QAM ($m = 6$), the total rate simplifies to $R_\mathrm{tot\text{-}4D} = 4R_{\mathrm{DM}} + 12R_\mathrm{C} - 8$, with $0 < R_{\mathrm{DM}} \le 2$ and $\frac{2}{3} \le R_\mathrm{C} < 1$.

{As illustrated in Fig.~\ref{Fig. System_FSO_PCS}, the proposed system leverages ideal CSI feedback to dynamically optimize the transmitted signal's probability distribution in response to changing channel conditions. Specifically, when the channel quality improves or degrades, instantaneous channel attenuation is calculated (Eq.~\eqref{Eq.A_inst}) and compared with the control range. When the attenuation depth is less than the control range (Eq.~\eqref{Eq.gamma_R}), the shaping parameter $L$ is correspondingly increased or decreased, thereby adjusting the distribution matching rate $R_{\mathrm{DM}}$ and, consequently, the total net bit rate $R_{\mathrm{tot\text{-}4D}}$, as governed by Eqs.~(12)–(16) and Eq.~\eqref{Eq.R_total}. This control mechanism exhibits a monotonic relationship: a change in $L$ directly alters $R_{\mathrm{tot\text{-}4D}}$, which in turn shifts the post-FEC SNR threshold $\gamma_{\mathrm{thr}}$ required for error-free transmission. The system is designed to select a threshold that is just below the instantaneous received SNR, thereby ensuring error-free communication while maximizing spectral efficiency. To enable real-time implementation, a lookup table (LUT) mapping $R_{\mathrm{tot\text{-}4D}}$ to the appropriate shaping parameter $L$ can be precomputed from the curves in Fig.~\ref{Fig. L_vs_SNR_thr}. Since the instantaneous achievable rate $\mathrm{GMI}_{\mathrm{4D}}$ is a monotonically increasing function of the received SNR (as detailed in Sec.~II.A of~\cite{farid2007outage}), each target net bit rate uniquely determines a required SNR threshold $\gamma_{\mathrm{thr}}$}
\begin{equation}
    \mathrm{\gamma_{thr}} = \mathrm{GMI}_{\mathrm{4D}}^{-1}(R_\mathrm{{tot \text{-} 4D}}).
    \label{eq. gamma_thr}
\end{equation}

By flexibly adjusting $R_{\mathrm{tot\text{-}4D}}$, a control range $\gamma_{\mathrm{R}}$ for the SNR threshold can be established to accommodate instantaneous channel fading. This range is defined as
\begin{equation}
\begin{split}
    \mathrm{\gamma_{R}} =  \frac{\mathrm{GMI}_{\mathrm{4D}}^{-1}(R_\mathrm{{tot \text{-} 4D}}^{\mathrm{max}})}{\mathrm{GMI}_{\mathrm{4D}}^{-1}(R_\mathrm{{tot \text{-} 4D}}^{\mathrm{min}}) } =  \frac{\mathrm{\gamma_{thr}^{max}}}{\mathrm{\gamma_{thr}^{min}}},
\end{split}
\label{Eq.gamma_R}
\end{equation}
where $\gamma_{\mathrm{R}}$ represents the maximum instantaneous attenuation range controllable by the A-PCS system. In our model, the channel gain follows the distribution given in Eq.~\eqref{eq:h}. The ideal channel gain $h_{\mathrm{ideal}}$, in the absence of turbulence and pointing errors, is expressed as
\begin{equation}
h_{\mathrm{ideal}} = A_0 h_l,
\label{Eq.h_ideal}
\end{equation}
where $A_0$ and $h_l$ are deterministic parameters. Substituting Eq.~\eqref{Eq.h_ideal} into Eq.~\eqref{eq:h}, the instantaneous attenuation $A_{\mathrm{inst}}$ relative to the ideal case is derived as
\begin{equation}
A_{\mathrm{inst}}(h) = \frac{\gamma_{\mathrm{FSO}}(h_{\mathrm{ideal}})}{\gamma_{\mathrm{FSO}}(h)} = \frac{A_0 h_l}{h},
\label{Eq.A_inst}
\end{equation}
when $A_{\mathrm{inst}} \leq \gamma_{\mathrm{R}}$, error-free transmission can be maintained through rate adaptation; thus, outage occurs only when $A_{\mathrm{inst}} > \gamma_{\mathrm{R}}$. Defining $h_0$ such that $A_{\mathrm{inst}}(h_0) = \gamma_{\mathrm{R}}$, we obtain $h_0 = \frac{A_0 h_l}{\gamma_{\mathrm{R}}}$. By substituting $h_0$ into Eq.~\eqref{eq:h}, The outage probability $P_{\mathrm{out}}$ for given turbulence and pointing error severities using~[34, Eq.(15)], denoted by $[\sigma_{\mathrm{R}}^2, \sigma_{\mathrm{s}}]$, is then
\begin{equation}
\begin{split}
{P_{\mathrm{out}}}&{(\sigma_{\mathrm{R}}^2, \sigma_{\mathrm{s}}, \gamma_{\mathrm{R}})  = \int_0^{h_0} f_h(h) \mathrm{d}h} \\& {= \frac{\gamma^2}{\Gamma(\alpha) \Gamma(\beta)}  \times  G_{2,4}^{3,1} \left[ \frac{\alpha \beta}{\mathrm{\gamma_{R}}} \,\middle|\begin{array}{c} 1, \gamma^2+1 \\ \gamma^2, \alpha, \beta, 0 \end{array}  \right].}
\end{split}
\label{Eq. P_out}
\end{equation}

\begin{figure*}[!ht]
\centering
\includegraphics[width=1\linewidth]{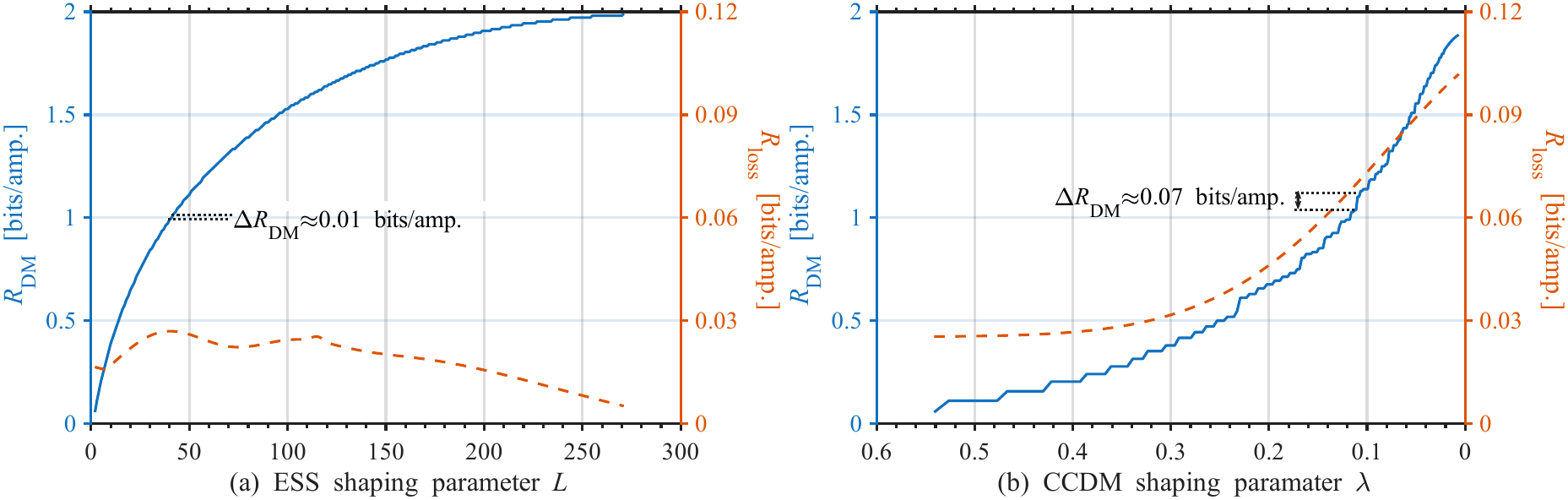}
\caption{
Comparison of net bit rate $R_{\mathrm{DM}}$ and rate loss $R_{\mathrm{loss}}$ per distribution matcher (DM) for ESS and CCDM under a fixed block length $n = 108$.
(a) Variation of $R_{\mathrm{DM}}$ with ESS shaping parameter $L$ {from Eq.~\eqref{Eq.L}}.
(b) Variation of $R_{\mathrm{DM}}$ with CCDM shaping parameter $\lambda$ {from Eq.~\eqref{Eq. lambda}.}
The A-PCS scheme based on ESS exhibits finer rate-regulation granularity and lower rate loss compared to CCDM.
}
\label{Fig. L_R_DM_R_loss}
\end{figure*}

For an A-PCS system, the maximum net bit rate $R_{\mathrm{tot\text{-}4D}}^{\mathrm{max}}$ is achieved when $h = h_{\mathrm{ideal}}$. Outage occurs when $A_{\mathrm{inst}} > \gamma_{\mathrm{R}}$, corresponding to $h < \frac{A_0 h_l}{\gamma_{\mathrm{R}}}$. The instantaneous transmission rate $R_{\mathrm{inst}}$ is adaptively adjusted based on real-time attenuation. For channel gains in the range $\frac{A_0 h_l}{\gamma_{\mathrm{R}}} \leq h \leq A_0 h_l$, $R_{\mathrm{inst}}$ is set to ensure $\gamma_{\mathrm{thr}} \leq \gamma_{\mathrm{FSO}}$, guaranteeing error-free transmission. When $h > A_0 h_l$, $R_{\mathrm{inst}}$ is fixed at $R_{\mathrm{tot\text{-}4D}}^{\mathrm{max}}$, which is 10 bits/4D in our design. Thus, $R_{\mathrm{inst}}$ is defined as
\begin{equation}
    R_{\mathrm{inst}} = \left\{\begin{array}{l} R_\mathrm{{tot \text{-} 4D}}^{\mathrm{max}},\quad \quad \quad \quad \quad \quad h > A_{\mathrm{0}}h_{\mathrm{l}}
\\ \mathrm{GMI}_{\mathrm{4D}}(\gamma_{\mathrm{FSO}}(h)),~~~~\frac{A_{\mathrm{0}}h_{\mathrm{l}}}{\mathrm{\gamma_{\mathrm{R}}}}\leq h \leq A_{\mathrm{0}}h_{\mathrm{l}}
\\ 0,(\text{Outage}) ~~\quad\quad\quad~~~ h < \frac{A_{\mathrm{0}}h_{\mathrm{l}}}{\mathrm{\gamma_{\mathrm{R}}}}
\end{array}\right.
\label{eq. R_inst}
\end{equation}
consequently, the ergodic capacity $C_{\mathrm{erg}}$ can be mathematically expressed as
\begin{equation}
\begin{split}
    C_{\mathrm{erg}} = & \int_{\frac{A_{\mathrm{0}}h_{\mathrm{l}}}{\mathrm{\gamma_{\mathrm{R}}}}}^{A_{\mathrm{0}}h_{\mathrm{l}}}\mathrm{GMI}_{\mathrm{4D}}(\gamma_{\mathrm{FSO}}(h))f(h)dh+ \\&\int_{A_{\mathrm{0}}h_{\mathrm{l}}}^{\infty }R_\mathrm{{tot \text{-} 4D}}^{\mathrm{max}}f(h)\mathrm{d}h,
    \end{split}
    \label{Eq.C_erg}
\end{equation}
where the integral can be computed numerically via Monte Carlo methods.

{In summary, the mathematical framework is built from three components: (i) the composite FSO channel model (Eq.~\eqref{eq:h}) linking SNR to channel state; (ii) the ESS DM, which yields the finite‑length rate loss and the 4D GMI (Eq.~\eqref{eq:GMI_DM}); and (iii) the adaptive control rule (Eq.~\eqref{eq. R_inst}) that sets the instantaneous rate based on channel attenuation. These expressions are directly evaluated to compute the two key performance parameters: the outage probability (Eq.~\eqref{Eq. P_out}) via integration of the channel PDF, and the ergodic capacity (Eq.~\eqref{Eq.C_erg}) via Monte Carlo expectation of the instantaneous rate over the non‑outage region. All numerical results are generated from these formulas.}

\subsection{A-PCS Performance Assessment}

ESS adjusts the maximum energy level via the shaping parameter $L$ in Eq.~\eqref{Eq.L}, thereby modifying the total number of valid sequences and tuning the net bit rate $R_{\mathrm{DM}}$. In contrast, CCDM directly generates amplitude sequences matching a target probability distribution via arithmetic coding~\cite{schulte2015constant}, allowing adjustment of $R_{\mathrm{DM}}$ through the shaping parameter $\lambda$ in Eq.~\eqref{Eq. lambda}. As shown in Fig.~\ref{Fig. L_R_DM_R_loss}, ESS achieves quasi-continuous control over $R_{\mathrm{DM}}$ by varying $L$ from $2$ to $271$. The resulting rate granularity $\Delta R_\mathrm{DM}$ is approximately $0.01$~bits/amp., covering a range from $0.05$ to $1.98$~bits/amp.. Throughout this range, the rate loss $R_{\mathrm{loss}}$ remains below $0.027$~bits/amp., with a minimum of $0.005$~bits/amp.. In comparison, CCDM-with $\lambda$ varying from $0.006$ to $0.54$-exhibits a coarser granularity of $\Delta R_\mathrm{DM} \approx 0.05\text{--}0.07$~bits/amp., which is about seven times larger than that of ESS. The achievable rate range for CCDM is $0.05$-$1.88$~bits/amp., slightly narrower than ESS. Moreover, as $\lambda$ increases, the rate loss of CCDM grows exponentially, reaching up to $0.1$~bits/amp. when $R_{\mathrm{DM}}$ lies between 1$\sim$2~bits/amp.. Such high rate loss severely constrains the achievable capacity.

\begin{figure}[!t]
\centering
\includegraphics[width=1\linewidth]{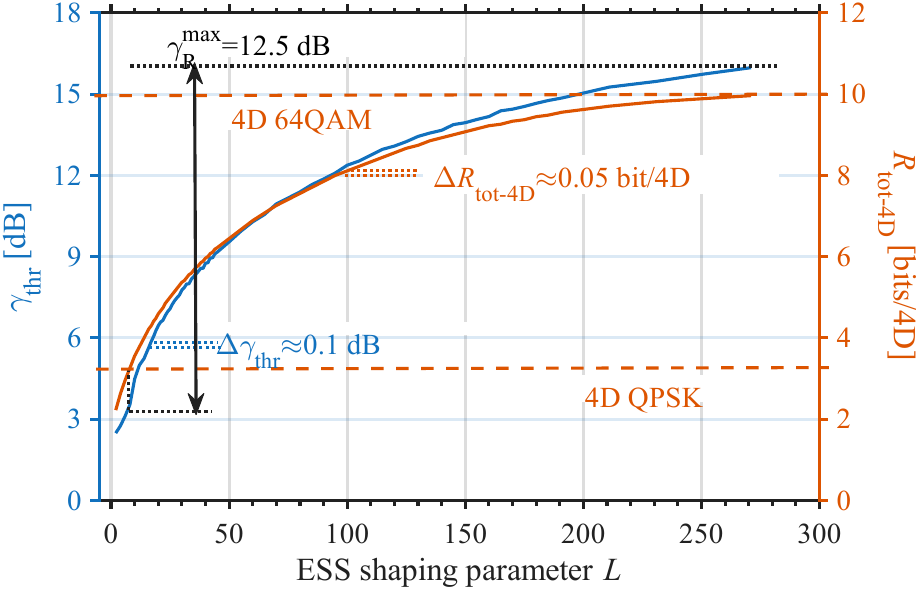}
\caption{
Performance of A-PCS-ESS with fixed block length $n = 108$ and FEC rate $R_{\mathrm{C}} = 5/6$.
Left: Post-FEC error-free transmission SNR threshold $\gamma_\mathrm{thr}$ versus ESS shaping parameter $L$. Right: Total net bit rate of 4D PCS-64QAM $R_\mathrm{tot\text{-}4D}$ versus $L$.
Orange dashed lines indicate $R_\mathrm{tot\text{-}4D}$ for conventional 4D-64QAM and 4D-QPSK.
A-PCS-ESS demonstrates wide dynamic range and fine granularity in rate adaptation.
}
\label{Fig. L_vs_SNR_thr}
\end{figure}

For each value of $R_{\mathrm{DM}}$ in Fig.~\ref{Fig. L_R_DM_R_loss}, the corresponding total net bit rate $R_\mathrm{tot\text{-}4D}$ is computed via Eq.~\eqref{Eq.R_total}. We maintain a fixed block length $n=108$ and an FEC rate of $5/6$. Under the sign-amplitude separation architecture shown in Fig.~\ref{Fig. System_FSO_PCS}, $R_{\mathrm{DM}}$ ranges from 2 to 10. The maximum $R_\mathrm{tot\text{-}4D}$ equals that of conventional 4D-64QAM, while the minimum falls below that of 4D-QPSK. This occurs because, under strong shaping, PCS-64QAM approaches QPSK behavior: most DM outputs correspond to low-energy amplitudes, yet each requires four amplitude encodings, reducing coding efficiency. Therefore, when $R_{\mathrm{DM}} = 10/3$, the system switches directly to 4D-QPSK rather than further reducing $R_{\mathrm{DM}}$. Consequently, $R_\mathrm{tot\text{-}4D}$ is regulated within the interval $[R_\mathrm{tot\text{-}4D}^\mathrm{min}, R_\mathrm{tot\text{-}4D}^\mathrm{max}] = [10/3, 10]$ (accounting for FEC overhead), as indicated by orange dashed lines in Fig.~\ref{Fig. L_vs_SNR_thr}. Within this range, the post-FEC error-free SNR threshold $\gamma_\mathrm{thr}$ varies from 3.5 dB to 16 dB, yielding a maximum A-PCS control range $\gamma_\mathrm{R}^\mathrm{max} = 12.5$ dB. The corresponding rate control granularity $\Delta R_\mathrm{tot\text{-}4D}$ is about 0.05~bits/4D, and the SNR control accuracy $\Delta\gamma_\mathrm{thr}$ is 0.1~dB. These results confirm that A-PCS-ESS enables quasi-continuous rate adaptation, closely approaching the theoretical capacity limit.

\begin{figure}[!t]
  \centering
  \includegraphics[width=1\linewidth]{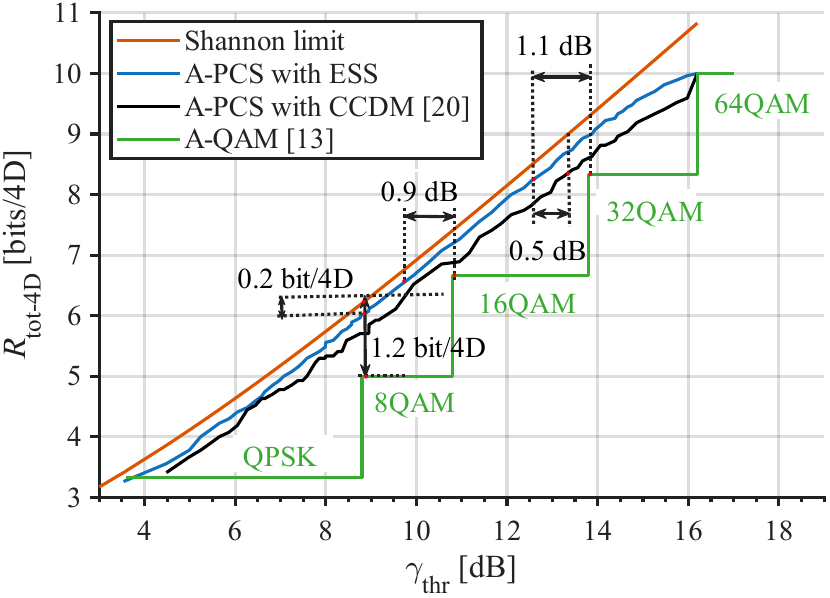}
\caption{Simulation of A-PCS-ESS/CCDM with a fixed block length $n = 108$ and $R_{\mathrm{C}} = 5/6$, the capacity of Shannon limit and A-QAM are also presented as baseline. Post-FEC error-free transmission SNR threshold $\gamma_\mathrm{{thr}}$ versus total net bit rate $R_\mathrm{{tot \text{-} 4D}}$ {using Eq.~\eqref{Eq.R_total}}. A-PCS based on ESS shows a smoother regulation and higher net bit rate than A-PCS based on CCDM over the entire range, and a much higher spectral efficiency than that of conventional $M$-QAM. The relationship between $\gamma_\mathrm{{thr}}$ and $R_\mathrm{{tot \text{-} 4D}}$ can be stored as a LUT for A-PCS control.}
    \label{Fig. SNR_thr_vs_Net_rate}
\end{figure}

Fig.~\ref{Fig. SNR_thr_vs_Net_rate} compares A-PCS using ESS (A-PCS-ESS) and CCDM (A-PCS-CCDM), alongside conventional $M$-QAM adaptive schemes (A-QAM) and the Shannon limit. A-QAM employs five discrete modulation formats (QPSK, 8QAM, 16QAM, 32QAM, 64QAM), leading to significant gaps in both rate and SNR adaptation and suboptimal spectral efficiency. In contrast, A-PCS-ESS offers two key advantages: (1) near-continuous rate and SNR control granularity improves spectrum utilization; (2) low rate loss and energy optimization via the Maxwell-Boltzmann distribution provide substantial shaping gains. Specifically, the gap to the Shannon limit is only about $0.2$~bits/4D-significantly smaller than the $1.2$~bits/4D gap of conventional 8QAM. Compared to uniform 16QAM, a shaping gain of $0.9$~dB is achieved; this is $1.1$~dB higher than uniform 32QAM and $0.5$~dB superior to A-PCS-CCDM, owing to lower rate loss. These results underscore the wide dynamic range, high spectral efficiency, and fine adaptation granularity of A-PCS-ESS. The mapping between $\gamma_\mathrm{thr}$ and $R_\mathrm{tot\text{-}4D}$ can be stored as a LUT for real-time adaptive control. 

\subsection{{Discussion on actual implementation and complexity}}

{The ESS DM operates with a fixed block length of $n = 108$ symbols, which is independent of the FEC codeword length (64,800 bits under the DVB-S2 standard with rate 5/6). Since the DM block and the FEC codeword are processed independently, the shaping parameter $L$ can be updated on a per-DM-block granularity if desired, offering finer temporal resolution than the codeword duration. More importantly, the atmospheric turbulence coherence time $\tau_\mathrm{c}$ typically ranges from 2 to 10 ms, which is orders of magnitude larger than the symbol duration (on the order of nanoseconds). Even at high symbol rates, the duration of a single DM block (108 symbols) remains well below $\tau_\mathrm{c}$, ensuring that the channel state is effectively constant over multiple consecutive DM blocks. Consequently, block-by-block adaptation is both feasible and sufficient for tracking channel variations. The adaptation itself is implemented via a simple lookup table (LUT) that maps the target net bit rate or estimated instantaneous SNR to the corresponding $L$ value, as illustrated in Fig.~\ref{Fig. System_FSO_PCS} and discussed around Eq.~\eqref{eq. gamma_thr}. The computational overhead of the LUT lookup and trellis reconfiguration is negligible compared to the DM block duration. These practical considerations confirm that the proposed A-PCS-ESS scheme is fully compatible with realistic framing and processing constraints without requiring fundamental architectural modifications.}

{In terms of complexity and storage requirements, arithmetic coding-based CCDM (AC-CCDM) only needs to store the distribution range and source statistics, resulting in lower storage demands. However, because AC-CCDM is a highly serial algorithm, it requires frequent division, multiplication, and comparisons involving large integer operations, which leads to significant latency in high-speed scenarios. In contrast, ESS requires only addition and subtraction operations, making the computation extremely simple and achieving very low latency. Furthermore, by employing bounded-precision ESS (BP-ESS), storage requirements can be significantly reduced~\cite{gultekin2020probabilistic}.}

{\textbf{Example 2} \textit{For a block length of $N = 108$, an amplitude set $\mathcal{A} = \{1,3,5,7\}$, and a shaping parameter $L \in [2, 271]$, the ESS trellis requires storing an $N \times L$ matrix (when using bounded‑precision arithmetic). For $L = 271$, this matrix contains up to $N \times L \approx 2.93 \times 10^4$ states.}}

{For Example 2, storing each state’s path count as a 32‑bit integer results in a memory footprint of roughly hundreds of kilobytes, which is easily accommodated by modern DSPs or FPGAs. Moreover, ESS encoding and decoding require only $k$ and $n$ recursive steps, respectively (a total serialism of $k + n$), which is significantly less computationally intensive than CCDM and the iterative processing of the subsequent LDPC decoder. Reconfiguration to adapt the shaping rate requires only updating the energy bound $E_{\text{max}}$ (equivalently $L$) and recomputing the trellis table—an operation that can be performed within the channel’s coherence time in adaptive transmission scenarios. Additionally, ESS offers finer granularity and lower rate loss, as demonstrated in the paper, while incurring only modest memory overhead. This makes ESS a practical choice for real‑time adaptive transmission over time‑varying channels.}


\section{Simulation Results under FSO Channel with Turbulence and Pointing Errors}\label{Simulation Results}

In this section, we evaluate the performance of the A-PCS system over an FSO channel impaired by turbulence and pointing errors. The turbulence severity is characterized by the Rytov variance $\sigma_\mathrm{R}^2$, as defined in Eq.~\eqref{eq:Rytov}. Similarly, the pointing error severity is quantified by the standard deviation of the pointing error displacement $\sigma_{\mathrm{s}}$ at the receiver, provided in Eq.~\eqref{eq:hp}.  The parameters for this numerical analysis, detailed in Table~\ref{Tab:2}, are adopted from prior work~\cite{farid2007outage,roa2025irradiance,aman2025effective}. We consider a point-to-point link with a span of $3$~km. The transmitter operates at a wavelength of $\lambda = 1550$~nm. The receiver is equipped with an aperture radius of $a = 5$~cm and an O/E conversion efficiency of $\kappa = 1$. For the channel parameters, we assume an attenuation of $0.2$~dB/km under clear air conditions, resulting in a path loss coefficient of $h_l = 0.73$. Regarding impairment parameters, $\sigma_{\mathrm{s}}$ ranges from $0.01$ (low pointing errors) to $0.6$ (high pointing errors), while $\sigma_\mathrm{R}^2$ varies between $0.01$ (weak turbulence) and $1.2$ (strong turbulence). {We consider both ideal CSI feedback—specifically, feedback without delay—and non-ideal CSI feedback—specifically, feedback with delay. In the non-ideal scenario, the coherence time of atmospheric turbulence $\mathrm{\tau_\mathrm{c}}$ varies between $2$ and $10$~ms~\cite{guiomar2022coherent,mostafa2012channel} and can be modeled using exponentially correlated processes via first-order stochastic differential equations~\cite{bykhovsky2016simple}, while the feedback delay $T_\mathrm{d}$ is set to range from $1$ to $5$~ms~~\cite{haque2023survey}.}



\begin{table}[t]
    \centering 
    \caption{FSO System Configuration}
    \label{Tab:2}
    \begin{tabular}{lc} 
        \toprule[1.5pt] 
        \multicolumn{2}{c}{\textbf{FSO link}} \\
        \midrule[0.75pt] 
        \textbf{Parameter} & \textbf{Value} \\
        \midrule[0.75pt]
        Wavelength ($\lambda$) & 1550 nm \\
        Efficiency of O/E conversion ($\kappa$) & 1 \\
        Tx-Rx Link distance (L) & 3 km \\
        Attenuation & 0.2 dB/km \\
        Receiver radius (a) & 5 cm \\
        Corresponding Beam radius at 3 km ($w_z$) & $\cong 2.1~\mathrm{m}$\\
        Rytov Variance ($\sigma_\mathrm{R} ^ 2$) & 0.01--1.2 \\ 
        Jitter standard deviation ($\sigma_\mathrm{s}$) & 0.01 m--0.6 m \\
        A-PCS range ($\mathrm{\gamma_{R}}$) & 12.5~dB \\
        Turbulence coherence time ($\mathrm{\tau_\mathrm{c}}$) & 2 ms--10 ms \\
        CSI feedback delay ($T_\mathrm{d}$) & 1 ms--5 ms \\
        \bottomrule[1.5pt] 
    \end{tabular}
\end{table}

\subsection{Ideal CSI feedback}

\begin{figure}[!t]
  \centering
  \includegraphics[width=1\linewidth]{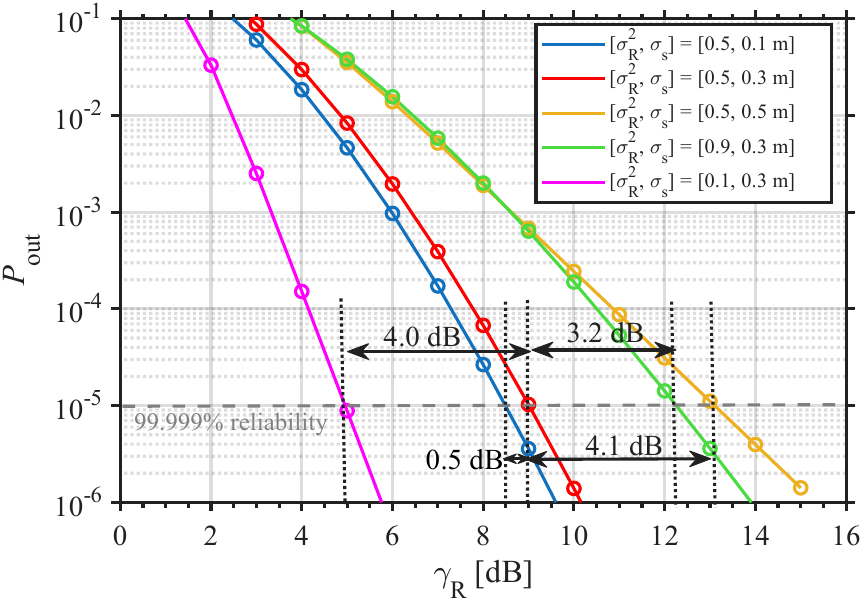}
\caption{{A-PCS control range $\mathrm{\gamma_{\mathrm{R}}}$ versus outage probability $P_{\mathrm{out}}$ with five different pairs of $[ \sigma_\mathrm{R}^2,\sigma_\mathrm{s}]$ which is indicated in legend. Solid lines represent results calculated using Eq.~\eqref{Eq. P_out}, while circles represent Monte Carlo simulation results. The outage probability of $10^{-5}$, which represents a reliability of 99.999$\%$, is shown as a gray dashed line as a baseline. An increase in pointing errors will rapidly increase the outage probability.}}
    \label{Fig. gamma_R_vs_Pout}
\end{figure}

Figure~\ref{Fig. gamma_R_vs_Pout} presents the variation of outage probability with respect to the control range for five distinct combinations of $\sigma_\mathrm{s}$ and $\sigma_\mathrm{R}^2$. The outage probability level of $10^{-5}$, which corresponds to a reliability of $99.999\%$~\cite{khalighi2014survey}, is depicted as a gray dashed line serving as a baseline. In the first scenario, $\sigma_\mathrm{R}^2$ is fixed at $0.5$, while $\sigma_\mathrm{s}$ is set to $0.1$~m, $0.3$~m, and $0.5$~m. As the pointing error increases, the minimum control range required to achieve the target outage probability of $10^{-5}$ rises. Specifically, for $\sigma_\mathrm{s} = 0.1$~m, a control range of $8.5$~dB is necessary. For $\sigma_\mathrm{s} = 0.3$~m, the requirement increases to $9$~dB, representing an increment of $0.5$~dB compared to the $0.1$~m case. When $\sigma_\mathrm{s}$ is increased to $0.5$~m, the transmission quality degrades significantly, necessitating an additional control range of $4.1$~dB to $13.1$~dB to maintain the same reliability. {Notably, the requirement of $13.1$~dB exceeds the A-PCS system's maximum control range of $12.5$~dB, implying that under such severe pointing errors, the system is unable to achieve its reliability goal through adaptive modulation adjustments, thereby resulting in an outage.} In the second scenario, $\sigma_\mathrm{s}$ is fixed at $0.3$~m, and $\sigma_\mathrm{R}^2$ is varied as $0.1$, $0.5$, and $0.9$. The corresponding control range requirements are: $5$~dB for $\sigma_\mathrm{R}^2 = 0.1$, $9.0$~dB for $\sigma_\mathrm{R}^2 = 0.5$ (indicating a degradation of $4.0$~dB), and an additional $3.2$~dB degradation to $12.2$~dB for $\sigma_\mathrm{R}^2 = 0.9$, approaching the system's operational limit. This behavior can be explained by the Rayleigh distribution of the pointing error. An increase in $\sigma_\mathrm{s}$ simultaneously amplifies the variance in both elevation and horizontal dimensions, leading to a rapid deterioration in transmission performance.

\begin{figure}[!t]
  \centering
  \includegraphics[width=1\linewidth]{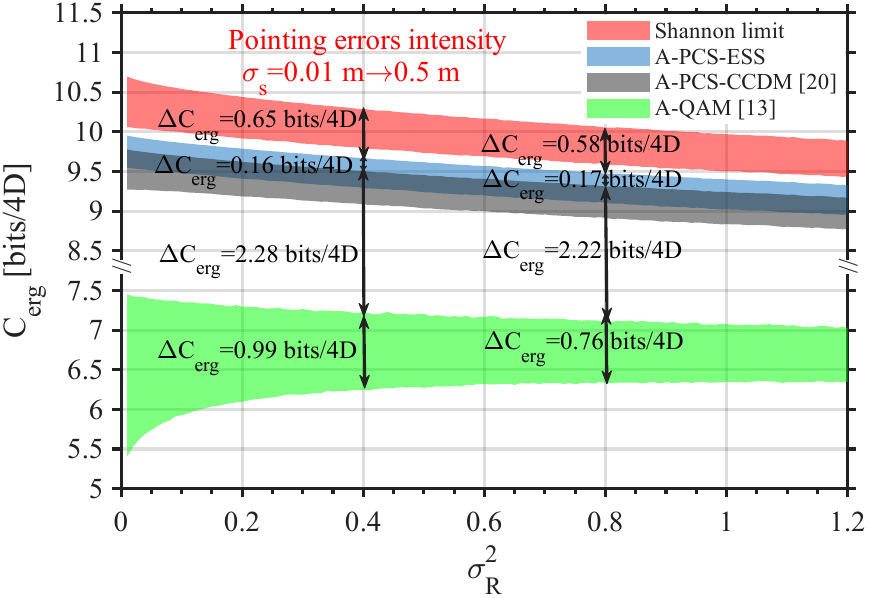}
\caption{{$C_{\mathrm{erg}}$ versus $\sigma_\mathrm{R}^2$ and $\sigma_\mathrm{s}$, calculated using Eq.~\eqref{Eq.C_erg}, with $12.5$~dB A-PCS range. The colored areas represent the performance changes when the pointing error intensity $\sigma_{\mathrm{s}}$ varies between $0.01$~m and $0.5$~m. A-PCS-ESS demonstrates superior performance across the entire range of channel variations. }}
    \label{Fig. Sigma_R_vs_C_erg}
\end{figure}

{To holistically evaluate the system performance from the perspective of average spectral efficiency, we introduce the ergodic capacity $C_{\mathrm{erg}}$ as a core metric, as defined in Eq.~\eqref{Eq.C_erg}. Unlike a point-to-point comparison, Fig.~\ref{Fig. Sigma_R_vs_C_erg} provides a global perspective by employing a colored contour map to visualize the distribution of $C_{\mathrm{erg}}$ across a two-dimensional parameter space defined by the pointing error severity $\sigma_{\mathrm{s}}$ (ranging from $0.01$ m to $0.5$ m) and the turbulence intensity $\sigma_{\mathrm{R}}^2$ (ranging from $0.01$ to $1.2$). This comprehensive visualization yields several key insights: First, across the entire parameter space, the performance surface of the A-PCS-ESS scheme consistently lies above that of A-PCS-CCDM~\cite{brandao2022cooperative}, maintaining a stable gap of approximately $0.17$ bits/4D. This advantage is directly attributable to the finer rate granularity and lower rate loss inherent to the ESS method, allowing it to match channel conditions more efficiently. Second, compared to the Shannon limit surface, both A-PCS schemes exhibit a persistent gap of approximately $0.58$–$0.65$ bits/4D. This disparity reveals the inherent capacity ceiling of the current PCS-64QAM modulation scheme, which is constrained by the rate loss induced by discrete coding and finite block length DM. Finally, in contrast to the surface of conventional adaptive QAM (A-QAM)~\cite{mai2014adaptive}, the colored region of the A-PCS schemes resides at a noticeably higher altitude, thereby providing intuitive confirmation of the significant advantage of probabilistic shaping in enhancing average capacity. Under the specific channel condition where $[\sigma_{\mathrm{R}}^2, \sigma_{\mathrm{s}}] = [0.4, 0.5\ \text{m}]$, the performance advantage of A-PCS-ESS is particularly pronounced, achieving a gain of up to $2.55$ bits/4D over conventional A-QAM. Furthermore, the finer rate control mechanism enhances the robustness of both A-PCS variants against pointing errors. Notably, A-PCS-ESS demonstrates more stable rate loss across the entire control range, as evidenced by its performance fluctuation consistently around $0.38$ bits/4D, which is lower than the $0.42$ bits/4D observed for A-PCS-CCDM. Meanwhile, it can be observed that the Shannon limit of ideal adaptive transmission varies by approximately $0.5$ bits/4D under the influence of pointing errors, a variation range slightly larger than that of the two A-PCS designs. This is because the two A-PCS designs experience outages when the fading depth exceeds the control range, while ideal adaptive transmission, unconstrained by such control mechanisms, does not. Consequently, the ideal scheme exhibits lower information entropy under severe channel deterioration, thereby reducing $C_{\mathrm{erg}}$ and expanding its overall variation range. In summary, Fig.~\ref{Fig. Sigma_R_vs_C_erg} serves as a comprehensive performance map, quantifying the tangible benefits of probabilistic shaping offered by the A-PCS designs and delineating the remaining gap to the theoretical optimum under diverse channel impairments.}

\begin{figure*}[!t]
  \centering
  \includegraphics[width=1\linewidth]{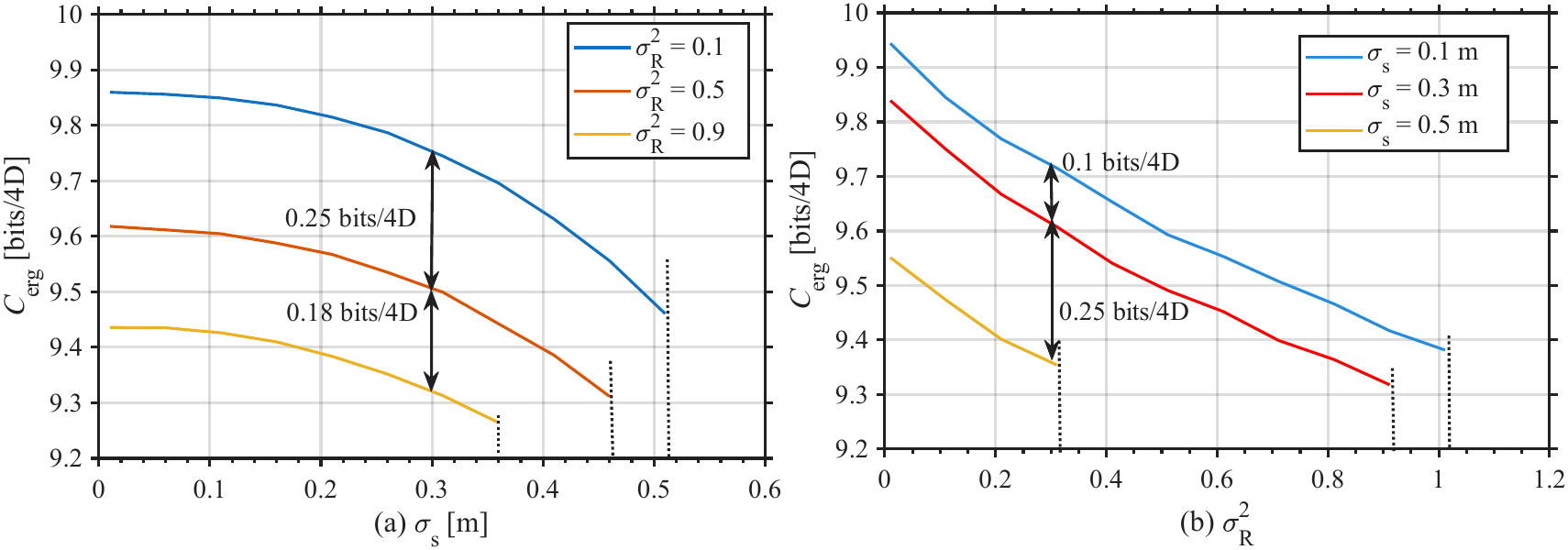}
\caption{{Performance of ergodic capacity $C_{\mathrm{erg}}$, calculated using Eq.~\eqref{Eq.C_erg}, in different pairs of $[\sigma_\mathrm{R}^2,\sigma_\mathrm{s}]$ with control range $\mathrm{\gamma_{\mathrm{R}}}$ is set to 12.5~dB as shown in Fig.~\ref{Fig. SNR_thr_vs_Net_rate}. $C_{\mathrm{erg}}$ will be null when $P_{\mathrm{out}}$ is greater than $10^{-5}$. Increasing $\sigma_\mathrm{s}$ and $\sigma_\mathrm{R}^2$ will lead to a continuous decrease in $C_{\mathrm{erg}}$.}}
    \label{Fig. sigma_R_2_s_vs_R}
\end{figure*}

To better compare the effects of atmospheric turbulence and pointing errors, as shown in Fig.~\ref{Fig. sigma_R_2_s_vs_R}, we fix either $ \sigma_\mathrm{R}^2 $ or $ \sigma_\mathrm{s} $ while varying the other parameter, with the control range $ \gamma_{\mathrm{R}} $ set to its maximum value of 12.5 dB. According to Fig.~\ref{Fig. SNR_thr_vs_Net_rate}, a transmission interruption is triggered when the control range required to maintain the target outage probability ($ 10^{-5} $) exceeds 12.5 dB. In the left sub-figure of Fig.~\ref{Fig. sigma_R_2_s_vs_R}, an increase in $ \sigma_\mathrm{s} $ intensifies the instantaneous attenuation, leading to a monotonic decrease in $ C_{\mathrm{erg}} $. For example, when $ \sigma_\mathrm{R}^2 = 0.5 $, $ C_{\mathrm{erg}} $ degrades steadily by $0.25$~bits/4D relative to the case with $ \sigma_\mathrm{R}^2 = 0.1 $. A further increase in $ \sigma_\mathrm{R}^2 $ to $0.9$ causes an additional degradation of $0.18$~bits/4D, and $ C_{\mathrm{erg}} $ continues to decline until the transmission interruption point. In the right sub-figure, $ C_{\mathrm{erg}} $ decreases as $ \gamma_{\mathrm{R}} $ increases. The case with $ \sigma_\mathrm{s} = 0.3 $ demonstrates greater stability than $ \sigma_\mathrm{s} = 0.1 $, exhibiting a deterioration of $0.1$~bits/4D. When $ \sigma_\mathrm{s} $ is increased to $0.5$, the degradation extends by $0.25$~bits/4D. Notably, at the transmission interruption point, $ C_{\mathrm{erg}} $ attains its maximum value for $ \sigma_\mathrm{s} = 0.5 $. This analysis reveals that although elevated $ \sigma_\mathrm{s} $ values lead to consistently low $ C_{\mathrm{erg}} $, the associated instantaneous decay fluctuations become excessively violent, resulting in a high incidence of out-of-control decays. As a consequence, the target interruption probability cannot be maintained before $ C_{\mathrm{erg}} $ drops to a critically low level.

\begin{figure}[!t]
  \centering
  \includegraphics[width=1\linewidth]{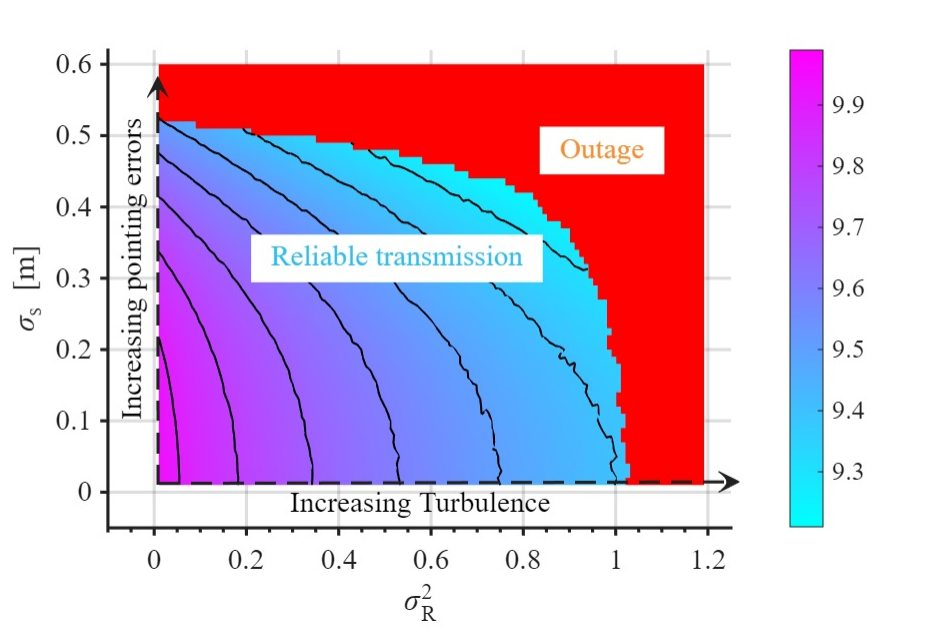}
\caption{{Performance of ergodic capacity $C_{\mathrm{erg}}$ using Eq.~\eqref{Eq.C_erg} in different pairs of $[\sigma_\mathrm{R}^2,\sigma_\mathrm{s}]$ with control range $\mathrm{\gamma_{\mathrm{R}}}$ is set to 12.5~dB. The heatmap depicts the value and trend of $C_{\mathrm{erg}}$, with the color bar on the right indicating the scale. Transmission is deemed reliable if $P_{\mathrm{out}} < 10^{-5}$. The proposed A-PCS system can maintain reliable transmission and high spectral efficiency over a wide range of channel variations.}}
    \label{Fig. R_ave_vs_outage}
\end{figure}

{Finally, in Fig.~\ref{Fig. R_ave_vs_outage}, we present the ergodic capacity $C_{\mathrm{erg}}$ as a function of $\sigma_\mathrm{s}$ and $\sigma_\mathrm{R}^2$ for a target outage probability of $10^{-5}$, with adaptive control applied over the full A-PCS range of $12.5$~dB. The heatmap offers a visualization of the global performance trends depicted in Fig.~\ref{Fig. Sigma_R_vs_C_erg}, with a particular emphasis on the operational boundaries and abrupt performance transitions. It contrasts the distinct failure modes induced by the two channel impairments. Along the vertical axis (varying $\sigma_\mathrm{s}$), the warm-colored (high-capacity) region rapidly contracts once $\sigma_\mathrm{s}$ exceeds approximately $0.4$~m, culminating in a complete outage (no data region) at $\sigma_\mathrm{s} \approx 0.51$~m. This behavior visually corroborates our earlier assertion that pointing errors precipitate a cliff-like drop in system performance . In contrast, along the horizontal axis (varying $\sigma_\mathrm{R}^2$), the transition from warm to cool colors is relatively gradual, indicating that $C_{\mathrm{erg}}$ declines steadily with increasing turbulence strength, with reliable transmission sustained up to $\sigma_\mathrm{R}^2 \approx 1.02$. A deeper interpretation of the heatmap's boundary data reveals a critical distinction: the performance degradation along the horizontal axis (turbulence) can reach lower $C_{\mathrm{erg}}$ values. This implies that while stronger turbulence introduces deeper instantaneous fades, their statistical probability increases only gradually, allowing the system to barely maintain communication through adaptive adjustments. Conversely, the outage along the vertical axis (pointing errors) occurs at a comparatively higher $C_{\mathrm{erg}}$ value, suggesting that an increase in $\sigma_\mathrm{s}$ sharply elevates the probability of deep fades, thereby more rapidly eroding the system's ability to sustain the target outage probability.}

\subsection{{Delayed CSI feedback}}

\begin{figure*}[!t]
\centering
\includegraphics[width=1\linewidth]{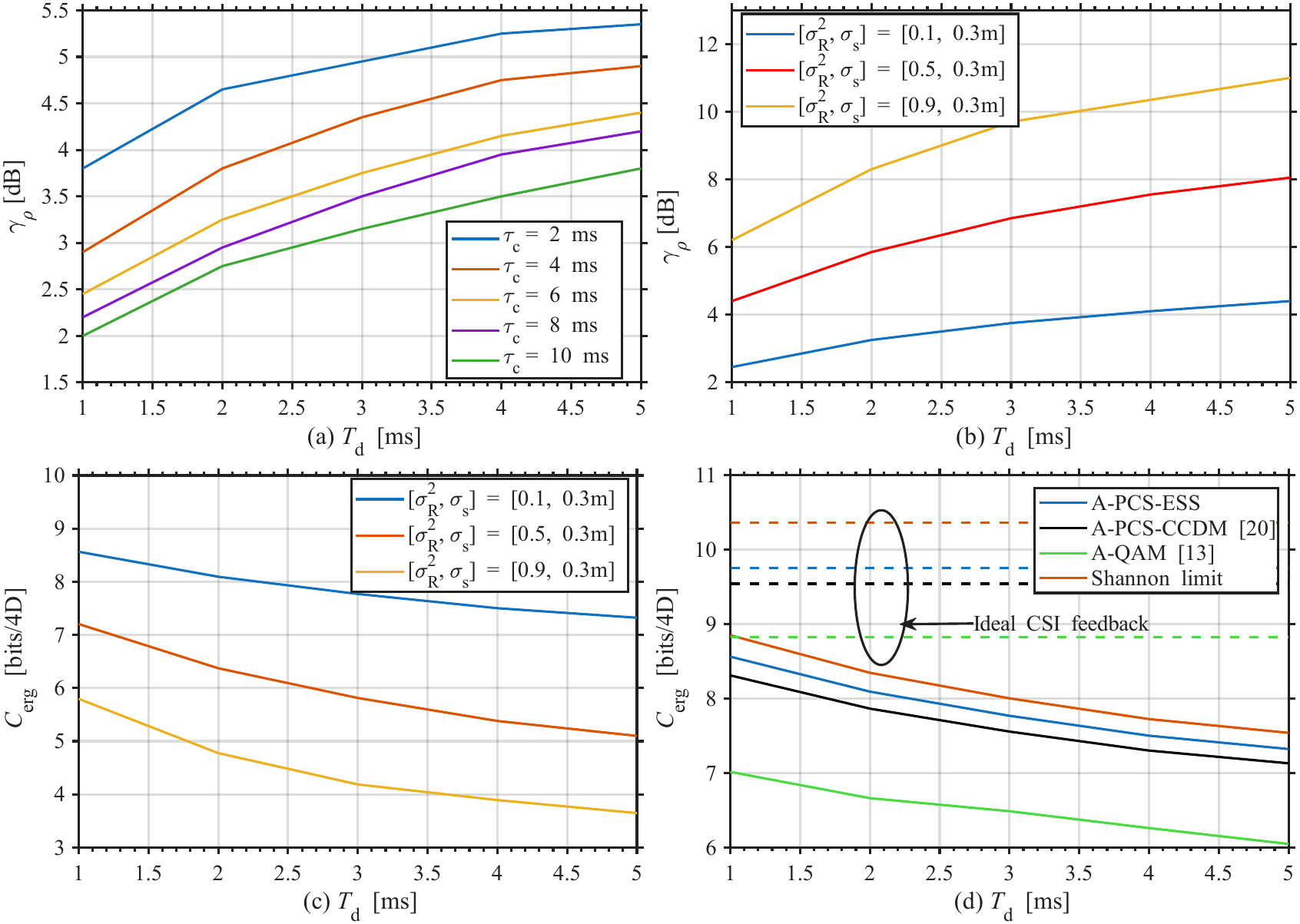}
\caption{{Comparison of SNR redundancy $\gamma_\mathrm{\rho}$ and ergodic capacity $C_{\mathrm{erg}}$ under different feedback delays $T_\mathrm{d}$ when $P_{\mathrm{out}} < 10^{-5}$. (a) Variation of $\gamma_\mathrm{\rho}$ versus $T_\mathrm{d}$ under different $\tau_\mathrm{c}$ with $[\sigma_{\mathrm{R}}^2, \sigma_{\mathrm{s}}] = [0.1, 0.3\ \text{m}]$. (b)  Variation of $\gamma_\mathrm{\rho}$ versus $T_\mathrm{d}$ under different pairs of turbulence intensity $\sigma_{\mathrm{R}}^2$ with $\tau_\mathrm{c} = 6~\mathrm{ms}$ and $\sigma_{\mathrm{s}} = 0.3~\text{m}$. (c) Variation of $C_{\mathrm{erg}}$ versus $T_\mathrm{d}$ under  different turbulence intensity $\sigma_{\mathrm{R}}^2$ with $\tau_\mathrm{c} = 6~\mathrm{ms}$ and $\sigma_{\mathrm{s}} = 0.3~\text{m}$. (d) Variation of $C_{\mathrm{erg}}$ versus $T_\mathrm{d}$ for different adapted schemes with $[\sigma_{\mathrm{R}}^2, \sigma_{\mathrm{s}}] = [0.1, 0.3\ \text{m}]$ and $\tau_\mathrm{c} = 6~\mathrm{ms}$. The performance of these four adaptive schemes under the ideal condition of zero-delay feedback is presented using dashed lines.}}
\label{Fig. Delay_vs_R}
\end{figure*}

{In practical systems, the delay $T_{\mathrm{d}}$ incurred by channel estimation, processing, and the feedback loop is unavoidable, rendering the CSI used by the transmitter outdated. Such non-ideal feedback can cause a severe mismatch between the preset modulation format and the instantaneous channel conditions, potentially leading to additional capacity loss or even an outage. To mitigate this issue, we propose a proactive defense strategy: a SNR redundancy, $\gamma_{\mathrm{\rho}}$, is added to the attenuation depth computed based on the delayed CSI. This redundancy is designed to provide a safety margin against potential channel degradation that may occur during the delay period. the system's control strategy is as follows:}
\begin{equation}
    {R_{\mathrm{inst}} = \left\{\begin{array}{l} R_\mathrm{{tot \text{-} 4D}}^{\mathrm{max}},\quad \quad \quad \quad \quad \quad h > A_{\mathrm{0}}h_{\mathrm{l}}\mathrm{\gamma_{\mathrm{\rho}}}
    \\ \mathrm{GMI}_{\mathrm{4D}}(\gamma_{\mathrm{FSO}}(\frac{h}{\mathrm{\gamma_{\mathrm{\rho}}}})),~~~\frac{A_{\mathrm{0}}h_{\mathrm{l}}\mathrm{\gamma_{\mathrm{\rho}}}}{\mathrm{\gamma_{\mathrm{R}}}}\leq h \leq A_{\mathrm{0}}h_{\mathrm{l}}\mathrm{\gamma_{\mathrm{\rho}}}
    \\ R_\mathrm{{tot \text{-} 4D}}^{\mathrm{min}} ~~\quad\quad\quad\quad\quad~~~ h < \frac{A_{\mathrm{0}}h_{\mathrm{l}}\mathrm{\gamma_{\mathrm{\rho}}}}{\mathrm{\gamma_{\mathrm{R}}}}
    \end{array}\right.}
\end{equation}
{where adaptive adjustment is performed based on the attenuation depth fed back—augmented with redundancy $\gamma_{\mathrm{redun}}$—to mitigate potential further channel degradation caused by delay. Furthermore, given that real-time CSI conditions may also improve following a delay, if the total equivalent attenuation depth—augmented by this redundancy—exceeds the maximum modulation depth of the A-PCS, the system does not immediately interrupt transmission. Instead, it autonomously switches to the minimum data rate, $R_{\text{tot-4D}}^{\text{min}}$, thereby prioritizing connectivity under extreme conditions. }

{Fig.\ref{Fig. Delay_vs_R}(a) quantitatively reveals the fundamental influence of channel temporal variability—characterized by the turbulence coherence time $\tau_{\mathrm{c}}$—on the tolerance to feedback delay. The family of curves demonstrates that for a given feedback delay $T_{\mathrm{d}}$, the required SNR redundancy $\gamma_{\mathrm{\rho}}$ decreases significantly as $\tau_{\mathrm{c}}$ increases. For example, when $T_{\mathrm{d}} = 5$ ms, a system with $\tau_{\mathrm{c}} = 10$ ms requires only about $3.8$~dB of redundancy, whereas a system with $\tau_{\mathrm{c}} = 2$ ms requires approximately $5.35$~dB.  $\tau_{\mathrm{c}}$ represents the correlation time of the channel state. The larger the $\tau_{\mathrm{c}}$, the higher the probability that the channel remains stable during the delay, and thus the lower the redundency(i.e., $\gamma_{\mathrm{\rho}}$) needed to hedge against uncertainty. Therefore in environments with slowly varying channels (large $\tau_{\mathrm{c}}$), a relatively longer feedback delay can be tolerated without incurring a prohibitive redundancy penalty. Fig.\ref{Fig. Delay_vs_R}(b) further investigates the impact of channel fading depth—governed by the turbulence intensity $\sigma_{\mathrm{R}}^2$—on the redundancy requirement. Under fixed $\tau_{\mathrm{c}}$ and $\sigma_{\mathrm{s}}$ conditions, stronger turbulence (larger $\sigma_{\mathrm{R}}^2$) necessitates higher SNR redundancy $\gamma_{\mathrm{\rho}}$ to combat the same feedback delay. Taking $T_{\mathrm{d}} = 5$~ms as an example, for a relatively mild channel of $[\sigma_{\mathrm{R}}^2, \sigma_{\mathrm{s}}] = [0.1, 0.3~\text{m}]$, the redundancy requirement is approximately $5$~dB. In contrast, for a strong turbulence channel of $[0.9, 0.3~\text{m}]$, this requirement surges to about $11$~dB. This data underscores the severity of the feedback delay problem in strong turbulence environments: when severe channel fading is compounded by outdated information, a considerable safety margin is required to maintain reliability. Since the control range of A-PCS-ESS spans $12.5$~dB, the system can maintain the target outage probability across varying turbulence intensities, coherence times, and feedback delays by adjusting the SNR redundancy.}


{The strategy of incorporating SNR redundancy to bolster reliability is not without its penalties; the direct and quantifiable cost is a reduction in the system’s ergodic capacity. Fig.\ref{Fig. Delay_vs_R}(c)  demonstrate that for a given turbulence condition (fixed $\sigma_{\mathrm{R}}^2$), the ergodic capacity $C_{\mathrm{erg}}$ monotonically declines as the feedback delay $T_{\mathrm{d}}$ increases. This is because a larger $T_{\mathrm{d}}$ necessitates a greater $\gamma_{\mathrm{\rho}}$ to maintain the target outage probability, effectively diverting SNR that could otherwise be used for higher-order modulation to enhance the data rate. Under strong turbulence (high $\sigma_{\mathrm{R}}^2$), the capacity starts from a lower baseline and exhibits a more pronounced decline with increasing delay. In the most adverse simulated scenario ($T_{\mathrm{d}} = 5$~ms, $[\sigma_{\mathrm{R}}^2, \sigma_{\mathrm{s}}] = [0.9, 0.3\ \text{m}]$), the ergodic capacity plummets to a mere $3.65$~bits/4D. Fig.\ref{Fig. Delay_vs_R}(d) offers a comprehensive comparison of the resilience of various adaptive schemes to feedback delay. By juxtaposing the performance curves under ideal (dashed lines) and non-ideal (solid lines) feedback, the vertical gap directly visualizes the capacity penalty incurred by delay. First, a finer modulation granularity enhances a scheme’s immunity to delay. Owing to its coarse modulation steps, conventional A-QAM suffers the most severe degradation, with a capacity drop of $1.81$~bits/4D at a mere $T_{\mathrm{d}} = 1$~ms (from $8.82$ to $7.01$~bits/4D). In contrast, the more finely granular A-PCS-ESS shows a markedly smaller decline of only $1.19$~bits/4D (from $9.75$ to $8.56$~bits/4D) under identical conditions. This disparity illustrates the superiority of probabilistic shaping in mitigating the effects of imperfect CSI. Second, the performance advantage of A-PCS-ESS over competing schemes is robust against delay. It consistently maintains a performance edge of approximately $0.2$~bits/4D over A-PCS-CCDM, thereby reaffirming the effectiveness of the ESS algorithm in achieving precise rate matching, even under non-ideal feedback conditions. It is worth noting that in the presence of feedback delay, the gap between A-PCS-ESS and the Shannon limit narrows further. This occurs because the maximum modulation rates for A-PCS-ESS, A-PCS-CCDM, and A-QAM are all capped at $10$~bits/4D (based on PCS-64QAM),, whereas the Shannon limit can theoretically reach higher values; therefore, by employing higher-order PCS schemes—such as PCS-256QAM—the A-PCS-ESS system could both mitigate the sensitivity to feedback delay and further approach the Shannon limit.}

\section{Discussion and Conclusion}
\label{Conclusion}

{We propose a comprehensive A-PCS architecture based on ESS, which explicitly accounts for the rate loss incurred in practical deployments due to finite-length distribution matching. Specifically, for FSO channels subjected to particular atmospheric conditions—including atmospheric turbulence, pointing errors, and path loss—we derive analytical expressions for transmission reliability and ergodic capacity under a given target outage probability constraint. Numerical simulation results demonstrate that, even under adverse conditions characterized by strong turbulence and severe pointing errors, the proposed scheme achieves a transmission reliability of up to 99.999\%. Compared to existing methods, this scheme offers finer-grained control and enhanced spectral efficiency. Furthermore, the results obtained under non-ideal feedback scenarios indicate that the system can adapt to a wide range of channel environments and feedback delays while maintaining its performance advantage over existing approaches. Future research will focus on extending the applicability of this scheme to even more challenging channel environments, incorporating accurate CSI prediction algorithms to facilitate reliability and capacity analyses, and experimentally validating the proposed architecture.}


\section*{Acknowledgments}
This work was supported by the Special Science Foundation of Shaanxi Provincial Department of Education (No. 25JK0679), National Natural Science Foundation of China (No. 62571176), Anhui Provincial Natural Science Foundation (No. 2508085J041), the Scientific Research Start-up Project of Xi'an University of Posts and Telecommunications (No. QD-LG2026-008), and the Natural Science Foundation of Shaanxi Provincial Department of Science and Technology (No. 2023-JC-JQ-58).

\bibliography{References}{}
\bibliographystyle{IEEEtran}

\end{document}